\documentclass[aps, pre, twocolumn,showpacs,amsmath, amssymb, superscriptaddress, reprint, longbibliography]{revtex4-1}
\usepackage[utf8]{inputenc}
\usepackage[T1]{fontenc}
\usepackage{xcolor,amsmath,amsfonts,amssymb,bm}
\usepackage{setspace}
\usepackage{tabularx}
\usepackage{array}
\usepackage{url}

\usepackage{graphicx}
\usepackage[textsize=footnotesize]{todonotes}

\definecolor{darkblue}{rgb}{0,0,0.6}
\definecolor{darkred}{rgb}{0.6,0,0}
\usepackage[colorlinks=true, urlcolor=darkblue, citecolor=darkblue, linkcolor=darkred, hyperfootnotes=false]{hyperref}

\newcommand{\iden}{{\bf 1}}
\newcommand{\ind}[1]{_{\mathrm{#1}}}

\renewcommand{\AA}{\boldsymbol{A}}
\newcommand{\BB}{\boldsymbol{B}}
\newcommand{\dd}{\mathrm{d}}

\newcommand{\ed}{\mathrm{e}}

\newcommand{\id}{\mathrm{i}}

\newcommand{\OO}{\boldsymbol{O}}

\newcommand{\qq}{\boldsymbol{q}}
\newcommand{\QQ}{\boldsymbol{Q}}
\newcommand{\rr}{\boldsymbol{r}}

\newcommand{\uu}{\boldsymbol{u}}

\newcommand{\eepsilon}{\boldsymbol{\epsilon}}

\newcommand{\ssigma}{\boldsymbol{\sigma}}
\newcommand{\uupsilon}{\boldsymbol{\upsilon}}

\DeclareMathOperator{\tr}{tr}

\begin{document}

\title{Elastic interactions in damage models of brittle failure}

\author{Vincent Démery}
\affiliation{Gulliver, CNRS, ESPCI Paris, PSL Research University, 10 rue Vauquelin, 75005 Paris, France}
\affiliation{Univ Lyon, ENS de Lyon, Univ Claude Bernard Lyon 1, CNRS, Laboratoire de Physique, F-69342 Lyon, France}

\author{Véronique Dansereau}
\affiliation{Univ. Grenoble Alpes, CNRS, ISTerre, 38000 Grenoble, France}

\author{Estelle Berthier}
\affiliation{Institut Jean Le Rond d’Alembert, UMR 7190, CNRS and Université Pierre et Marie Curie, 75005 Paris, France}
\affiliation{Department of Physics, North Carolina State University, Raleigh, North Carolina, USA}

\author{Laurent Ponson}
\affiliation{Institut Jean Le Rond d’Alembert, UMR 7190, CNRS and Université Pierre et Marie Curie, 75005 Paris, France}

\author{Jérôme Weiss}
\affiliation{Univ. Grenoble Alpes, CNRS, ISTerre, 38000 Grenoble, France}

\begin{abstract}
The failure of brittle solids involves, before macroscopic rupture, power-law distributed avalanches of local rupture events whereby microcracks nucleate and grow, which are also observed in for an elastic interface evolving in a non-homogeneous medium.  
For this reason, it is tempting to relate failure to the depinning of an elastic interface.
Here we compute the elastic kernel of the interface representing the damage field of a brittle solid.
In the case of a damage model of rupture under compression, which implements the Mohr-Coulomb criterion at the local scale, we show that the elastic kernel is unstable, and hence is very different from the kernels of usual interfaces.
We show that the unstable modes are responsible for the localization of damage along a macroscopic fault observed in numerical simulations.
At low disorder, the most unstable mode gives the orientation of the macroscopic fault that we measure in numerical simulations.
The orientation of the fault changes when the level of disorder is increased, suggesting a complex interplay of the unstable modes and the disorder.
\end{abstract}

\maketitle

\section{Introduction}\label{sec:intro}

The shear or compressive brittle failure of materials such as rock, concrete or ice, involves complex precursory phenomena of microcrack nucleation and growth. 
These local rupture events can be recorded with acoustic emission, that gives access to both their intensity and their location~\cite{Peng1972, Lockner1991, Baud2004, Fortin2006}.
At the onset of damage, these events are randomly distributed in the sample. But when approaching failure, they progressively concentrate along the plane of the future macroscopic fault~\cite{Lockner1991}.
%Their intensities are power-law distributed, with a cut-off that diverges as macroscopic failure is approached~\cite{Davidsen2007}.
Their intensities are power-law distributed~\cite{Scholz1968, Davidsen2007}, with an exponent that has been claimed to decrease as approaching final failure~\cite{Lei2004}. However, a recent re-interpretation argues for a power-law distribution with a constant exponent but a cut-off that diverges towards failure~\cite{Amitrano2012, Berthier2016}.
A similar phenomenology is encountered in the depinning transition of elastic interfaces~\cite{Fisher1998}, which has been used to investigate the effect of the size of a sample on its failure strength~\cite{Weiss2014}.
%This similarity has been used to draw a parallel between the two phenomena in the context of investigating the effect of the size of a sample on its failure strength~\cite{Weiss2014}.

The behavior of a material during brittle failure is qualitatively captured by damage models~\cite{Tang1997, Zapperi1997, Lyakhovsky1997, Amitrano1999, Girard2010}.
In these models, a damage variable $d(\rr,t)$ defines the local state of the material, from sane ($d=0$) to completely damaged ($d=1$). 
%The level of damage affects the elastic properties of the material; hence, a damage increase 
The level of damage can be seen as the ``height'' of an elastic interface. 
Since the elastic properties of the material depend locally on the damage, a rupture event occuring at the point $\rr$ generates a stress redistribution that may trigger rupture events at other points $\rr'$ in the sample.
This mechanism is responsible for the ``elasticity'' of the interface, which is encoded in the elastic kernel $\Psi(\rr)$: if damage increases by an amount $\delta d$ at the point $\rr$, the damage driving force at the point $\rr'$ increases by $\Psi(\rr'-\rr)\delta d$.

The shape of the elastic kernel deeply affects the properties of the interface, notably the critical exponents of the distributions of the size or duration of avalanches~\cite{Fisher1998, Kardar1998}.
In the case of classical elastic interfaces, the elastic kernel satisfies two properties: (i) it is \textit{negative}, i.e., it has only non-positive eigenvalues and the interface is stable; (ii) it is \textit{convex}, which means that an increase of the height of the interface at some point can only trigger an increase of the height at another point.
As a result, there are only two independent exponents, which define the universality class of the interface.

Recently, an anologous mapping to the depinning of an elastic interface has been discussed for the yielding transition of amorphous solids~\cite{Lin2014, Lin2014b, Nicolas2017}.
In this situation, the kernel contains soft modes and is still negative, but not convex: the stress redistribution generated by a plastic event can therefore be negative at some locations. 
This non-convexity results in the existence of a third independent exponent~\cite{Lin2014b}.

Hence, in order to transfer the results on elastic interfaces to the rupture of brittle solids, the elastic kernel and, in particular, both its convexity and negativity, should be determined.
The elastic kernel has already been computed for a damage model that implements the Mohr-Coulomb (MC) failure criterion locally~\cite{Weiss2014}. 
However, while it is clear that it is not convex, its eigenvalues have not been computed.

Here, we compute the elastic kernel for a general damage model in Fourier space, which allows a direct identification of its eigenmodes and eigenvalues.
In the case of a two-dimensional damage model with MC failure criterion, which has been used extensively to represent the compressive failure of rocks and ice~\cite[e.g.,][]{Amitrano1999, Girard2010}, we show that the elastic kernel is not negative, namely it has unstable modes. 
As a consequence, the corresponding interface model may not belong to the universality classes mentionned above, in particular, the number of critical exponents may be different.

The most direct consequence of the unstability of the elastic kernel is that the roughness of the interface representing the damage field may diverge.
The shape of the interface in this process can be investigated through the unstable modes.
As for the yielding of amorphous solids, in Fourier space, the elastic kernel depends only on the orientation of the wavevector, but not on its magnitude. 
Thus, along the most unstable direction, all modes diverge at the same rate.
For these reasons, we interpret the instability of the kernel as being at the origin of the damage localization observed in damage models.
Doing so, we relate the most unstable direction to the orientation of the macroscopic fault.

For damage models that use the MC failure criterion, our prediction differs from the angle of the plane that maximizes the Coulomb stress,
%\begin{equation}
%\theta\ind{MC}=\frac{\pi}{4}-\frac{\phi}{2},
%\end{equation}
%where $\phi$ is the angle of internal friction, 
which has been thought to give the orientation of conjugate faults both in the laboratory and at geophysical scales \citep{Schulson2004} and/or of macroscopic faults with respect to tectonic forces \citep{Reches1987}.
Notably, our prediction depends on the Poisson's ratio of the material and on the confinement applied on the lateral sides of the sample, which is not the case in the MC theory.

Our prediction is compared to numerical simulations, and an excellent agreement is found when the sample is initialized with a single, evanescent inclusion in the initial field of cohesive strength of the material.
When more disorder is introduced in the sample, the observed angle of localization deviates from our prediction but remains far from the MC angle.
The prediction of the orientation of the macroscopic fault and its implications in geoscience is the focus of Ref.~\cite{short}.

Finally, we show that localization also occurs when disorder is introduced via the initial Young's modulus instead of the cohesion.
In the context of an elastic interface, this is analogous to having an initial configuration of the interface that is not flat, but that evolves in an homogeneous medium. 
That the interface is unstable in this situation is at odds with interfaces with a negative kernel, which become flat in the absence of disorder.

This article is organized as follows.
The general damage model is introduced in Sec.~\ref{sec:model}.
The elastic kernel is computed in Sec.~\ref{sec:lin_stab} with a linear stability analysis.
The general analysis is applied to the planar model of damage failure under compression in Sec.~\ref{sec:model_compression}, and the prediction of the localization angle is given.
The numerical implementation of the model is detailed in Sec.~\ref{sec:num_model} and the results are presented in Sec.~\ref{sec:num_results}.
We conclude in Sec.~\ref{sec:conclusion}.

\section{Progressive damage model}\label{sec:model}

We consider a typical, simple progressive damage model, structured as follows.
The level of damage in a small region around $\rr$ is given by $d(\rr)\in[0,1]$, with $d=0$ for the intact material and $d=1$ for a completely broken material.
Locally, the level of damage affects the elastic properties of the material: its Young's modulus $E(d)$ and Poisson's ratio $\nu(d)$.
The exact form of $E(d)$ and $\nu(d)$ varies between models~\cite{Ponte_Castaneda1995, Amitrano1999, Berthier2017}.

At equilibrium, the stress field $\ssigma(\rr)$ in the sample obeys the momentum equation 
\begin{equation}\label{eq:div_sigma}
\nabla\cdot\ssigma(\rr) = 0.
\end{equation}
The stress field is related to the strain tensor $\eepsilon(\rr)$ through Hooke's law:
\begin{equation}
\ssigma = 2G\eepsilon + \lambda \tr(\eepsilon)\iden,
\label{eq:hooke}
\end{equation}
where the Lamé parameters are given by
\begin{align}
G & = \frac{E}{2(1+\nu)},\label{eq:lame_G}\\
\lambda & = \frac{E\nu}{(1+\nu)(1-2\nu)}. \label{eq:lame_lambda}
\end{align}
Note that Hooke's law (Eq.~(\ref{eq:hooke})) is local and relates, at every point of the sample, the quantities $\ssigma(\rr)$, $\eepsilon(\rr)$, $G(\rr)$ and $\lambda(\rr)$.

Another ingredient of progressive damage models is a local criterion for the evolution of the damage.
Usually, the criterion compares a driving force and a resistance; the damage increases if their difference $Y$ (the total force) is positive.
Here, we assume that the total force depends only on the local stress $\ssigma(\rr)$; it would be straightforward to introduce an additional dependence on the local elastic coefficients or on the local level of damage.
% \footnote{In such model, it is straightforward to introduce an additional dependence of the Lamé parameters on the local level of damage.}.
Assuming a linear relation between the total force and the damage increase, we can write
\begin{equation}
\alpha\frac{\partial d}{\partial t}(\rr,t) = \max(Y(\ssigma(\rr)), 0),
\end{equation}
where the parameter $\alpha$ can be absorbed in a redefinition of the time, so that we omit in the following.
%When the damage increases at $\rr$, the total force decreases at this point, so that 
In the quasistatic limit the damage evolution ensures that $Y\leq 0$ everywhere in the sample.

From these ingredients, a closed relation can be written for the evolution of damage, in the form
\begin{equation}
\frac{\partial d}{\partial t}(\rr,t) = F[d(\cdot,t), \ssigma^0](\rr),
\end{equation}
where $\ssigma^0$ is the external stress imposed on the sample.
We use brackets to underline the fact that the evolution of damage at a point $\rr$ depends on the level of damage everywhere in the sample.
This non-locality comes from the stress redistribution generated by a point where the damage, and thus the elastic coefficients, are different from their values elsewhere in the material.

\section{Linear stability analysis of the damage evolution}\label{sec:lin_stab}

In this section, we perform a linear stability analysis of the general damage model introduced above.
First, we linearize the elastic coefficients around a value $d^0$ of the damage, and the total force $Y$ around the external stress $\ssigma^0$ (Sec.~\ref{sub:elast_coeffs_total_force}).
Second, we compute the stress redistribution generated by a weak heterogeneity in the elastic coefficients, in the spirit of the stress redistribution around an ellipsoidal inclusion~\cite{Eshelby1957} (Sec.~\ref{sub:redistribution}).
Finally, we put these ingredients together to write the effect of the damage on the driving force as a convolution of the damage heterogeneities by the so-called \textit{elastic kernel} (Sec.~\ref{sub:kernel}).
The final result is given in Eqs.~(\ref{eq:kernel}, \ref{eq:damage_evol_linearized}).

\subsection{Elastic coefficients and damage criterion}\label{sub:elast_coeffs_total_force}

To perform the linear stability analysis, we first linearize all of the equations above.
The starting point is to assume that the damage fluctuations are weak, hence $d(\rr)=d^0+d^1(\rr)$, with $d^0$ the average damage in the sample and $d^1(\rr)\ll d^0$.

First, we linearize the Lamé parameters:
\begin{align}
G(d^0+d^1(\rr)) &\simeq G(d^0)+G'(d^0)d^1(\rr) = G^0[1+g^1(\rr)], \label{eq:taylor_G}\\
\lambda(d^0+d^1(\rr)) &\simeq \lambda(d^0)+\lambda'(d^0)d^1(\rr) = G^0[\ell^0+\ell^1(\rr)], \label{eq:taylor_lambda}
\end{align}
where we have defined
\begin{align}
G^0 & = G(d^0),\\
\ell^0 & = \frac{\lambda(d^0)}{G(d^0)},\\
g^1(\rr) & = \frac{G'(d^0)}{G(d^0)} d^1(\rr),\label{eq:def_g1}\\
\ell^1(\rr) & = \frac{\lambda'(d^0)}{G(d^0)} d^1(\rr). \label{eq:def_l1}
\end{align}

If the stress fluctuations $\ssigma^1(\rr)$ are weak compared to the external stress $\ssigma^0$, the total force, $Y(\ssigma)$, entering the damage criterion can then be expanded as follows, using a symmetric tensor $\uupsilon$:
\begin{equation}\label{eq:taylor_total_force}
Y(\ssigma^0+\ssigma^1(\rr))\simeq Y(\ssigma^0) + \uupsilon:	\ssigma^1(\rr).
\end{equation}
The colon in this expression denotes the contraction, $\uupsilon:\ssigma^1=\upsilon_{ij}\sigma^1_{ij}$, where a summation over the repeated indices is assumed.

Finally, the fluctuations of the stress field, $\ssigma^1(\rr)$, can be computed from the fluctuations of the elastic coefficients, $g^1(\rr)$ and $\ell^1(\rr)$. This is done in the next section.

\subsection{Stress redistribution due to heterogeneous elastic coefficients}\label{sub:redistribution}

If a sample with weakly heterogeneous Lamé parameters is submitted to an ``external stress'' $\ssigma^0$, then the stress in the sample is weakly heterogeneous: $\ssigma(\rr)=\ssigma^0+\ssigma^1(\rr)$, with $\|\ssigma^1(\rr)\|\ll\|\ssigma^0\|$.
Here, we compute the fluctuations $\ssigma^1(\rr)$ as a function of the fluctuations in the elastic coefficients, $g^1(\rr)$ and $\ell^1(\rr)$ (Eqs.~(\ref{eq:taylor_G}, \ref{eq:taylor_lambda})), assuming a domain of infinite dimensions.
This calculation is reminiscent of the work of Eshelby~\cite{Eshelby1957} and has been done with a slightly different approach in Ref.~\cite{Weiss2014}. For the sake of consistency, we present all steps of the calculation in the following.

We use the same notation for the stress $\ssigma(\rr)$, the strain $\eepsilon(\rr)$, and the displacement $\uu(\rr)$. The quantity with exponent $0$ refers to the zeroth order term that would be the solution for a homogeneous sample, and the quantity with exponent $1$ refers to the (small) fluctuations. 

We expand the elasticity equations to the first order in small quantities - discarding second order terms of the form $g^1(\rr) u^1(\rr)$ - and write it in components form as follow~:
\begin{align}
\partial_i\sigma_{ij}^1 & = 0, \label{eq:equil_o1}\\
\frac{\sigma_{ij}^1}{G^0} & = 2g^1\epsilon^0_{ij}+\ell^1\epsilon^0_{kk}\delta_{ij} + \partial_iu_j^1+\partial_j u_i^1+\ell^0\partial_k u_k^1\delta_{ij},\label{eq:hooke_o1}
\end{align}
where we drop the argument $\rr$ of the different fields to simplify the writing and use the definition of the strain as a function of the displacement for the order 1: $\epsilon_{ij}^1 = (\partial_i u_j^1+\partial_j u_i^1)/2$. Summation over repeated indices is assumed.

In order to solve for the displacement $u^1_i$, we substitute for $\sigma_{ij}^1$ ~(\ref{eq:hooke_o1}) in Eq.~(\ref{eq:equil_o1})~ and obtain
\begin{equation}
\partial_i\partial_iu^1_j+\left(1+\ell^0\right)\partial_j\partial_i u_i^1 = -2(\partial_j g^1)\epsilon^0_{ij}-(\partial_j\ell^1)\epsilon^0_{ii}.
\end{equation}
This equation can be solved in Fourier space, by introducing a function $f(\rr)$,
\begin{equation}
\tilde f(\qq)=\int f(\rr)\ed^{-\id\qq\cdot\rr}\dd\rr.
\end{equation}
The derivatives then become $\partial_i\to\id q_i$ and we get
\begin{equation}\label{eq:displacement_o1}
q^2\tilde u_j^1+\left(1+\ell^0 \right)q_j q_i \tilde u_i^1 = 2\id\tilde g^1 q_i\epsilon^0_{ij}+\id\tilde\ell^1 q_j\epsilon^0_{ii}.
\end{equation}

We can find the displacements from Eq.~(\ref{eq:displacement_o1}). 
The first step is to determine the divergence, $q_j\tilde u_j$, by multipliying Eq.~(\ref{eq:displacement_o1}) by $q_j$:
\begin{equation}
q^2(2+\ell^0)q_j\tilde u_j^1 = 2\id\tilde g^1q_iq_j\epsilon_{ij}^0+\id\tilde\ell^1 q^2\epsilon^0_{ii},
\end{equation}
hence
\begin{equation}
q_j\tilde u_j^1 = \frac{\id}{2+\ell^0}\left(2\tilde g^1\frac{q_iq_j}{q^2}\epsilon^0_{ij}+\tilde\ell^1\epsilon^0_{ii} \right).
\end{equation}
Using this expression in Eq.~(\ref{eq:displacement_o1}), we get
%\begin{equation}\label{eq:sol_displacement}
%\tilde u_j^1 = 2\id\tilde g^1 \frac{q_i}{q^2}\epsilon^0_{ij}+\id \frac{q_j}{q^2}\left(-2 \frac{1+\ell^0}{2+\ell^0}\tilde g^1\frac{q_i q_k}{q^2}\epsilon^0_{ik}+\frac{1}{2+\ell^0}\tilde\ell^1\epsilon^0_{ii} \right).
%\end{equation}
\begin{multline}\label{eq:sol_displacement}
\tilde u_j^1 = 2\id\tilde g^1 \frac{q_i}{q^2}\epsilon^0_{ij}\\+\id \frac{q_j}{q^2}\left(-2 \frac{1+\ell^0}{2+\ell^0}\tilde g^1\frac{q_i q_k}{q^2}\epsilon^0_{ik}+\frac{1}{2+\ell^0}\tilde\ell^1\epsilon^0_{ii} \right).
\end{multline}

\begin{widetext}

Now, we can insert Eq.~(\ref{eq:sol_displacement}) in Eq.~(\ref{eq:hooke_o1}) to get the stress redistribution in Fourier space:
\begin{align}
\frac{\tilde\sigma_{ij}^1}{G_0} & = 2\tilde g^1\epsilon^0_{ij}+\tilde\ell^1\epsilon^0_{kk}\delta_{ij} 
+ \id (q_i\tilde u_j^1+q_j \tilde u_i^1)+\id\ell^0 q_k \tilde u_k^1\delta_{ij}\\
& = 2\tilde g^1 \left(\epsilon_{ij}^0-\frac{q_iq_k\epsilon^0_{kj}+q_jq_k\epsilon^0_{ki}}{q^2}
+\frac{1}{2+\ell^0}\frac{q_kq_l}{q^2}\epsilon^0_{kl}\left[2(1+\ell^0)\frac{q_iq_j}{q^2}-\ell^0\delta_{ij} \right] \right) %\nonumber\\
%& \qquad 
+\tilde\ell^1 \frac{2}{2+\ell^0}\epsilon^0_{kk} \left(\delta_{ij}-\frac{q_iq_j}{q^2} \right).\label{eq:sol_stress_strain}
\end{align}
We can rewrite this expression in tensorial form using the tensor
\begin{equation}\label{eq:def_Q}
Q_{ij}(\qq)=\frac{q_iq_j}{q^2},
\end{equation}
and the contraction $[\AA\cdot\BB]_{ij} = A_{ik}B_{kj}$, leading to
\begin{equation}
\frac{\tilde\ssigma^1}{G_0} = 2\tilde g^1 \left(\eepsilon^0-\QQ\cdot\eepsilon^0-\eepsilon^0\cdot\QQ
+\frac{1}{2+\ell^0}\QQ:\eepsilon^0\left[2(1+\ell^0)\QQ-\ell^0\iden \right] \right) 
 +\tilde\ell^1 \frac{2}{2+\ell^0}\epsilon^0_{kk} \left(\iden-\QQ \right).
\end{equation}

The final step consists in expressing the stress redistribution, $\tilde\ssigma^1$, as a function of the external uniform stress, $\ssigma^0$ (i.e., instead of $\eepsilon^0$).
To do so, we invert Hooke's law (\ref{eq:hooke})
\begin{equation}
\eepsilon^0 = \frac{1}{2G^0} \left[\ssigma^0-\frac{\ell^0}{2+3\ell^0}\tr(\ssigma^0)\iden \right]
\end{equation}
and insert it into Eq.~(\ref{eq:sol_stress_strain}) using a tensorial notation. Using $\QQ:\iden = 1$, we obtain:
\begin{multline}
\tilde\ssigma^1 = \tilde g^1 \left(\ssigma^0-\QQ\cdot\ssigma^0-\ssigma^0\cdot\QQ+\frac{1}{2+\ell^0}(\QQ:\ssigma^0)\left[2(1+\ell^0)\QQ-\ell^0\iden \right] \right)
\\+(\tilde\ell^1-\ell^0\tilde g^1) \frac{2}{(2+\ell^0)(2+D\ell^0)}(\iden:\ssigma^0)(\iden-\QQ).
\end{multline}
This expression can be further simplified by defining the Oseen tensor
\begin{equation}\label{eq:def_oseen}
\OO(\qq)=\iden-\QQ(\qq), 
\end{equation}
and by using the fact that $(\QQ:\ssigma^0)\QQ=\QQ\cdot\ssigma^0\cdot\QQ$. We then obtain
\begin{equation}\label{eq:stress_redistribution}
\tilde\ssigma^1 = \tilde g^1 \left(\OO\cdot\ssigma^0\cdot\OO-\frac{\ell^0}{2+\ell^0}[(\iden-\OO):\ssigma^0]\OO\right)
+(\tilde\ell^1-\ell^0\tilde g^1) \frac{2}{(2+\ell^0)(2+3\ell^0)}(\iden:\ssigma^0)\OO.
\end{equation}
We note that $\qq\cdot\OO=0$, which implies $\qq\cdot\tilde\ssigma^1=0$. Hence the equilibrium condition is satisfied.

\subsection{Elastic kernel and stability analysis}\label{sub:kernel}

Combining the results of the two previous subsections (Eqs.~(\ref{eq:def_g1}, \ref{eq:def_l1}, \ref{eq:taylor_total_force}, \ref{eq:stress_redistribution})), we can write the fluctuation of the total force $Y^1(\rr)=Y(\ssigma(\rr))-Y(\ssigma^0)$, in Fourier space, as
\begin{align}
\tilde Y^1(\qq) & \simeq \uupsilon:\tilde\ssigma^1(\qq)\\
& = \tilde g^1(\qq) \uupsilon:\left(\OO\cdot\ssigma^0\cdot\OO-\frac{\ell^0}{2+\ell^0}[(\iden-\OO):\ssigma^0]\OO\right) +\left[\tilde\ell^1(\qq)-\ell^0\tilde g^1(\qq)\right] \frac{2}{(2+\ell^0)(2+3\ell^0)}(\iden:\ssigma^0)\uupsilon:\OO\\
& = \tilde \Psi(\qq) d^1(\qq).\label{eq:kernel_damage_fourier}
\end{align}
We have defined the elastic kernel 
\begin{multline}\label{eq:kernel}
\tilde \Psi(\qq) = \frac{G'(d^0)}{G(d^0)} \uupsilon:\left(\OO\cdot\ssigma^0\cdot\OO-\frac{\ell^0}{2+\ell^0}[(\iden-\OO):\ssigma^0]\OO\right)\\
+\left[\frac{\lambda'(d^0)}{G(d^0)}-\frac{\lambda(d^0)G'(d^0)}{G(d^0)^2}\right] \frac{2}{(2+\ell^0)(2+3\ell^0)}(\iden:\ssigma^0)\uupsilon:\OO,
\end{multline}
\end{widetext}
where the dependence of the Oseen tensor $\OO(\qq)$ (Eq.~(\ref{eq:def_oseen})) on $\qq$ has been omitted.
In real space, the product in Eq.~(\ref{eq:kernel_damage_fourier}) becomes a convolution product: $Y^1(\rr) = \Psi*d^1(\rr)$.
For instance, if the heterogeneity is a defect localized at $\rr=0$, meaning that $d^1(\rr)=d^1 \delta(\rr)$, the damage driving force at the point $\rr$ is $Y(\ssigma_0)+d^1\Psi(\rr)$. It also means that a localized increase of damage in $\bf{r} = 0$ is accompanied by a variation $\sim \Psi(\bf{r})$ of the driving force. This long-range redistribution mechanism may explain the bursts of failure events observed prior localization~\citep{Peng1972, Lockner1991, Baud2004, Fortin2006}.
Finally, the linearized law for the evolution of damage is
\begin{equation}\label{eq:damage_evol_linearized}
\frac{\partial d}{\partial t}(\rr,t) \simeq \max\left(Y(\ssigma^0)+\Psi*d^1(\rr,t), 0\right).
\end{equation}

With the elastic kernel at hand, we can perform the stability analysis. 
%Because the maximum in the evolution of damage (Eq.~(\ref{eq:damage_evol_linearized})) ensures that $d$ never decreases, the evolution of damage cannot be expanded around $d=0$. 
%Thus, we cannot perform a true linear stability analysis of $d$.
%We therefore focus on the stability of the elastic kernel $\Psi(\rr)$ itself.

Consider a perturbation of the damage field of the form $d^1(\rr)=d^1\cos(\qq\cdot\rr)$. It will induce a perturbation in the driving force, given by $Y^1(\rr)=\tilde\Psi(\qq)\cos(\qq\cdot\rr)$.
The perturbation vanishes if the driving force is smaller where the damage is larger, which is the case if $\tilde\Psi(\qq)<0$.
On the contrary, if $\tilde\Psi(\qq)>0$, the driving force is larger where the damage is larger and the perturbation will grow, leading to damage localization.

In Fourier space, the elastic kernel depends on the mode $\qq$ through the Oseen tensor $\OO(\qq)$. 
It thus depends only on the direction $\hat\qq$ of the mode, and not on its magnitude.
So if some mode $\qq$ is unstable, all the modes with the same direction $\hat\qq$ are also unstable: there is no preferred wavelength for the instability.
The localization observed in numerical models may thus corresponds to the localization along planes orthogonal to $\hat\qq$. 

%As a consequence, there is no preferred wavelength and the damage can localize along planes orthogonal to $\hat\qq$.
%, corresponding to the localization observed in nature and numerical models.

\section{Application to a planar model of damage failure under compression}\label{sec:model_compression}

%The expression (\ref{eq:stress_redistribution}) for the stress redistribution is heavy, and so is the expression of the kernel $\Psi$ (Eq.~(\ref{eq:kernel_gen}).
We apply the general expressions obtained in the previous section to a model of damage failure under compression, inspired from the models used in Refs.~\cite{Tang1997, Lyakhovsky1997, Zapperi1997, Amitrano1999}.

\subsection{Elastic coefficients}\label{sub:elastic_coeffs_exp}

In the models of Ref.~\cite{Zapperi1997} and Ref.~\cite{Amitrano1999}, the Young's modulus decays with damage as $E(d)=(1-d)E_0$ and Poisson's ratio is a constant, $\nu_0$, independant of damage.
Hence, the Lamé parameters (Eqs.~(\ref{eq:lame_G}, \ref{eq:lame_lambda})) are given by $G(d)=(1-d)G_0$ and $\lambda(d)=(1-d)\lambda_0$, and their derivatives are
$G'(d)=-G_0$, $\lambda'(d)=-\lambda_0$.
Using these expressions in Eq.~(\ref{eq:kernel}), we get
%\begin{equation}\label{eq:elastic_kernel_1}
%\tilde\Psi(\qq) = \frac{-1}{1-d^0} \uupsilon:  \left(\OO\cdot\ssigma^0\cdot\OO-\frac{\ell^0}{2+\ell^0}[(\iden-\OO):\ssigma^0]\OO\right).
%\end{equation}
\begin{multline}
\label{eq:elastic_kernel_1}
\tilde\Psi(\qq) = \\\frac{-1}{1-d^0} \uupsilon:  \left(\OO\cdot\ssigma^0\cdot\OO-\frac{\ell^0}{2+\ell^0}[(\iden-\OO):\ssigma^0]\OO\right).
\end{multline}

\subsection{External stress and damage criterion}\label{}

The stress applied on the sample is given by
\begin{equation}
\ssigma^0 =- \begin{pmatrix}
\sigma_1 & 0 & 0 \\ 0 & \sigma_2 & 0 \\ 0 & 0 & \sigma_3
\end{pmatrix},
\end{equation}
with $\sigma_1>\sigma_2>\sigma_3$ (these values are positive for compression). 
The direction 1 is the direction of maximum principal stress.

We use the Mohr-Coulomb criterion for damage. In this case, the difference between the driving force and the resistance is given by
\begin{equation}\label{eq:mohr_coulomb}
Y(\ssigma^0)=\sigma_1-\sigma_3-(\sigma_1+\sigma_3)\sin(\phi)-2\tau\ind{c}\cos(\phi),
\end{equation}
where $\phi = \tan^{-1}(\mu)$ is the angle of internal friction and $\mu$ is the internal friction coefficient.
In general, the values $\sigma_i$ denote the eigenvalues of $-\ssigma^0$. 

If the stress becomes $\ssigma=\ssigma^0+\ssigma^1$, then to the first order in $\ssigma^1$, the eigenvalues of $-\ssigma$ are given by $\sigma_i - \sigma^1_{ii}$, $1\leq i\leq 3$.
The Mohr-Coulomb criterion can thus be expanded:
%\begin{equation}
%Y(\ssigma^0+\ssigma^1) \simeq Y(\ssigma^0) - \sigma_{11}^1 + \sigma_{33}^1 + ( \sigma_{11}^1  +  \sigma_{33}^1 )\sin(\phi)= Y(\ssigma^0) + \uupsilon:\ssigma^1,
%\end{equation}
\begin{align}
Y(\ssigma^0+\ssigma^1) &\simeq Y(\ssigma^0) - \sigma_{11}^1 + \sigma_{33}^1 + ( \sigma_{11}^1  +  \sigma_{33}^1 )\sin(\phi) \nonumber\\
&= Y(\ssigma^0) + \uupsilon:\ssigma^1,
\end{align}
with
\begin{equation}\label{eq:upsilon_tensor}
\uupsilon = \begin{pmatrix}
-1+\sin(\phi) & 0 & 0 \\
0 & 0 & 0 \\
0 & 0 & 1+\sin(\phi)
\end{pmatrix}.
\end{equation}
This tensor enters the expression of the elastic kernel (Eq.~(\ref{eq:elastic_kernel_1})). 
% and $\sigma_1>\sigma_2>\sigma_3$ are the eigenvalues of $-\ssigma$ (they are positive for compression).

\subsection{Planar configuration}\label{}

Most of the numerical damage models are two-dimensional. Here we write more explicit formulas in this case.

The elastic coefficients to use in Eq.~(\ref{eq:elastic_kernel_1}) are the two-dimensionnal elastic coefficients, which are different for plane stresses and plane strains. 
For plane stresses,
\begin{align}
\ell^0 & = \frac{2\nu}{1-\nu},\\
\frac{\ell^0}{2+\ell^0} & = \nu.
\end{align}
For plane strains,
\begin{align}
\ell^0 & = \frac{2\nu}{1-2\nu},\\
\frac{\ell^0}{2+\ell^0} & = \frac{\nu}{1-\nu}.
\end{align}

The external stress is defined as
\begin{equation}
\ssigma^0=-\sigma_1 \begin{pmatrix}
1 & 0 \\ 0 & R
\end{pmatrix},
\end{equation}
where $\sigma_1>0$ and $R$ is the confinement ratio.
Considering two-dimensional, plane stresses conditions ($\sigma_3 = 0$) and a local MC failure criterion, failure may only occur if $\sigma_1 - q \sigma_2 \geq 0$, where $q = \left[ (\mu^2+1)^{1/2} + \mu \right]^2$ is the slope of the MC envelope. 
This is represented by the dotted lines that radiate from the origin in Fig. \ref{dcriterion}. 
Hence, for failure to occur, the confinement ratio should satisfy
\begin{equation}\label{eq:R_max}
R\leq R\ind{max} = q^{-1} 
= \frac{(\mu^2 + 1)^{1/2} - \mu}{(\mu^2 + 1)^{1/2} + \mu}
= \frac{1-\sin(\phi)}{1+\sin(\phi)}.
\end{equation}

The tensor $\uupsilon$ (Eq.~(\ref{eq:upsilon_tensor})) involved in the failure criterion is now given by
\begin{equation}
\uupsilon = \begin{pmatrix}
-1+\sin(\phi) & 0 \\
0 & 1+\sin(\phi)
\end{pmatrix}.
\end{equation}

The orientation of a wavevector $\qq$ is given by the polar angle $\omega$. 
With this notation, the Oseen tensor reads
\begin{equation}
\OO(\omega)= \begin{pmatrix}
\sin(\omega)^2 & -\sin(\omega)\cos(\omega) \\ -\sin(\omega)\cos(\omega) & \cos(\omega)^2
\end{pmatrix}.
\end{equation}
As a consequence, the kernel given in Eq.~(\ref{eq:kernel}) also depends only on the angle, $\omega$, in Fourier space.
Because of the symmetry of the problem, $\tilde \Psi(\omega)=\tilde \Psi(-\omega)$ and $\tilde \Psi(\omega)=\tilde \Psi(\pi-\omega)$.

The localization may occur along lines perpendicular to the most unstable mode, $\omega^*$, with an angle $\theta\ind{LS}$ with respect to the direction of maximum principal stress. 
Taking $\omega^*\in[0,\pi/2]$, the mode $\pi-\omega^*$ is as unstable as the mode $\omega^*$, and the failure perpendicular to it has an angle $\theta\ind{LS}=\pi-\omega^*-\pi/2=(\pi/2)-\omega^*\in[0,\pi/2]$.

\subsection{Elastic kernel and localization angle}\label{}

The kernel given in Eq.~(\ref{eq:kernel}) reads
%\begin{equation}\label{eq:kernel_2d}
%\tilde \Psi(\omega)=-\frac{2\sigma_1}{1-d^0} (1-R) \left(1+\frac{\ell^0}{2+\ell^0} \right)
%\left[\sin(\omega)^2-\frac{1+\sin(\phi)}{2} \right]
%\times \left[\sin(\omega)^2-\delta\right],
%\end{equation}
\begin{multline}\label{eq:kernel_2d}
\tilde \Psi(\omega)=-\frac{2\sigma_1}{1-d^0} (1-R) \left(1+\frac{\ell^0}{2+\ell^0} \right)\\\times
\left[\sin(\omega)^2-\frac{1+\sin(\phi)}{2} \right]
\times \left[\sin(\omega)^2-\delta\right],
\end{multline}
where
\begin{equation}
\delta=\frac{\frac{\ell^0}{2+\ell^0}-R}{\left(1+\frac{\ell^0}{2+\ell^0} \right)(1-R)}.
\end{equation}
We give $\delta$ for the plane stress and plane strain configurations:
\begin{align}
\delta_\text{plane stress} & = \frac{\nu-R}{(1+\nu)(1-R)},\\
\delta_\text{plane strain} & = \frac{\nu-(1-\nu)R}{1-R}.
\end{align}
In general, the elastic kernel could depend on damage, it does not for this particular model. Instead, Eq.~\ref{eq:kernel_2d} depends on the homogeneous $d^0$.

The kernel is maximal and positive for 
\begin{equation}
\sin(\omega^*)^2=\frac{1+\sin(\phi)+2\delta}{4}.
\end{equation}
The kernel is thus unstable, and localization may occur along lines with an angle 
\begin{equation}\label{eq:loc_angle}
\theta\ind{LS}=\arccos\left(\frac{\sqrt{1+\sin(\phi)+2\delta}}{2}\right)
\end{equation}
with respect to the direction of maximum principal stress (direction 1), which in the following, we refer to as the localization angle predicted by the linear stability analysis.

%\todo[inline]{Say somewhere that a damage independent kernel is a particular case.}

\subsection{Elastic kernel in real space}\label{}

In this section, we also give the expression for the kernel of the planar model, $\Psi(\rr)$ (Eq.~(\ref{eq:kernel_2d})), in real space. This expression was presented by Ref.~\cite{Weiss2014}. However, here, we provide a more explicit form and use it to determine the direction of maximal redistribution.

We start from the expression in Fourier space, Eq.~(\ref{eq:kernel_2d}), which we write in the more compact form
\begin{equation}
\tilde \Psi(\omega) = \alpha\left[\beta-\sin(\omega)^2\right]\left[\sin(\omega)^2-\delta\right],
\end{equation}
where $\alpha = 2\sigma_1 (1-R) \left(1+\frac{\ell^0}{2+\ell^0} \right)/(1-d^0)$ and $\beta = [1+\sin(\phi)]/2$.
Reintroducing the wavevector $\qq=(q_x, q_y)$, we rewrite
\begin{align}
\tilde \Psi(\qq) &= \alpha\left(\beta-\frac{q_y^2}{q^2}\right)\left(\frac{q_y^2}{q^2}-\delta\right)\\
& = \alpha \frac{-\beta\delta q_x^4+(\beta+\delta-2\beta\delta)q_x^2q_y^2+(\beta-1)(1-\delta)q_y^4}{q^4}.
%& = \alpha \frac{Aq_x^4+Bq_x^2 q_y^2+Cq_y^4}{q^4},
\end{align}
The three terms appearing in this expression can be Fourier-transformed individually. Their Fourier transforms are, up to a singular part proportionnal to $\delta(\rr)$:
\begin{align}
\frac{q_x^4}{q^4} & \to \frac{-x^4-6x^2y^2+3y^4}{8\pi r^6},\\
\frac{q_x^2 q_y^2}{q^4} & \to \frac{-x^4+6x^2y^2-y^4}{8\pi r^6}, \\
\frac{q_y^4}{q^4} & \to \frac{3x^4-6x^2y^2-y^4}{8\pi r^6}.
\end{align}
We thus get the kernel in real space
\begin{multline}
\Psi(x,y)=\frac{\alpha}{8\pi r^6}\left[(-3+2\beta+2\delta-\beta\delta)x^4\right.
\\\left.+6x^2y^2+(1-2\beta-2\delta)y^4 \right],
\end{multline}
which we can write in polar coordinates as
%\begin{equation}
%\Psi(r, \theta)=\frac{\alpha}{8\pi r^2}\left[(-3+2\beta+2\delta-\beta\delta)\cos(\theta)^4+6\cos(\theta)^2\sin(\theta)^2+(1-2\beta-2\delta)\sin(\theta)^4 \right].
%\end{equation}
\begin{multline}
\Psi(r, \theta)=\frac{\alpha}{8\pi r^2}\left[(-3+2\beta+2\delta-\beta\delta)\cos(\theta)^4\right.
\\\left.+6\cos(\theta)^2\sin(\theta)^2+(1-2\beta-2\delta)\sin(\theta)^4 \right].
\end{multline}

The direction, $\theta\ind{max}$, where this kernel is maximal for a given $r$ is given by 
%\begin{equation}\label{eq:theta_max}
%\theta\ind{max} = \arccos \left(\frac{\sqrt{1+\beta+\delta}}{2} \right) = \arccos \left(\sqrt{\frac{3+\sin(\phi)+2\delta}{8}} \right).
%\end{equation}
\begin{align}
\theta\ind{max} & = \arccos \left(\frac{\sqrt{1+\beta+\delta}}{2} \right) \nonumber\\
&= \arccos \left(\sqrt{\frac{3+\sin(\phi)+2\delta}{8}} \right). \label{eq:theta_max}
\end{align}
The angle where the redistribution is maximal is thus different from the angle of the most unstable mode, $\theta\ind{LS}$ (Eq.~\ref{eq:loc_angle}).

Two different angles have also been extracted from the stress redistribution due to local transformations in a deformed granular medium~\cite{Karimi2017}; these angles have been shown to arise in experiments at different stages of deformation~\cite{LeBouil2014}.

\section{Numerical simulations}
\label{sec:num_model}

\subsection{Damage model}\label{model}

We compare both the prediction from the most unstable mode of the elastic kernel, $\theta\ind{LS}$, and from the direction of the maximal stress redistribution, $\theta\ind{max}$, to the angle of localization of damage in a finite element based, progressive damage model. 

The model is two-dimensional and solves the momentum and constitutive equations presented in Sec.~\ref{sec:model}, Eqs.~(\ref{eq:div_sigma}) and (\ref{eq:hooke}). Plane stress conditions are assumed, hence the Lamé parameters are given by
\begin{align}
G & = \frac{E}{2(1+\nu)},\\
\lambda & = \frac{E\nu}{(1-\nu^2)}.
\end{align}

\begin{figure}[t]
\begin{center}
\includegraphics[width=8.5cm]{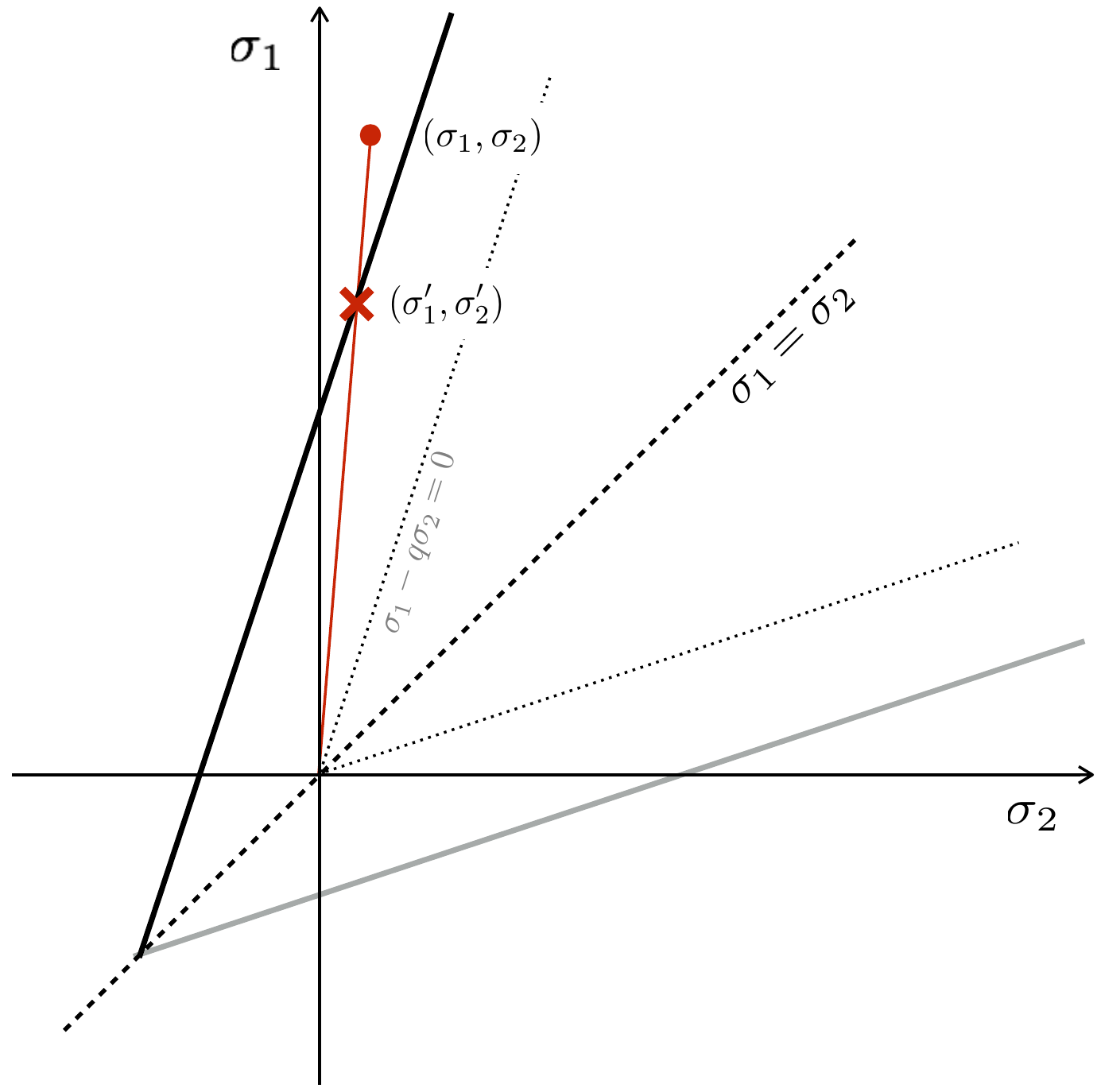}
\end{center}

\caption{MC damage criterion in the principal stresses plane (solid line). In the simulations performed here, the MC criterion is extended to tensile stresses and no truncation is used to close the envelope towards biaxial compression. The calculation of the distance to the damage criterion $d\ind{crit}$, defined by the intersection ($\sigma_1'$, $\sigma_2'$) of the line relating the state of stress ($\sigma_1$, $\sigma_2$) of a given element to the origin of the principal stress plane, is represented in red.} 
\label{dcriterion}

\end{figure}

The Mohr-Coulomb failure criterion is implemented at the local scale, that is, the scale of the model element. It is extented to tensile stresses (Fig.~\ref{dcriterion}). % the exact form the failure criteria in tension does not impact the results presented here. 

The level of damage is represented by a dimensionless scalar variable, $d$, the value of which varies between 0 and 1 as defined in Section ~\ref{sec:model}. The dependence of $E$ on $d$ is given by
\begin{align*}
E(d) & = E^0 (1-d),
\end{align*}
where $E^0$ is the elastic modulus of an undamaged element and Poisson's ratio is independant of $d$ (Sec.~\ref{sub:elastic_coeffs_exp}). 

The model is iterated in time. At each time step, the local state of stress over each model element is compared to the critical stress set by the MC failure criterion. Over-critical elements become damaged, which implies an increase in the local value of $d$ (hence a decrease in the local value of $E$). The increment in $d$ over any given damaged element is calculated so that to bring the state of stress of this element back on the failure envelope. 
%The assumption behind this choice is that over-critical states of stress are unphysical, because failure occurs before the material can support them. 
An assumption made while calculating this increment is that the deformation of the damaged element is conserved during the damage event, i.e., at initiation, the first effect of damage is to initiate a stress redistribution between neighbouring elements, which modifies the local state of stress, not strains. The post-damage state of stress is then given by the intersection of the MC envelope and of the line connecting the pre-damage state of stress with the origin, in the principal stress plane (Fig. \ref{dcriterion}). The increment in $d$ is:
\begin{align}
d' - d &= (1-d)\left(1- \frac{\sigma_{1}'}{\sigma_{1}}\right) = (1-d)\left(1- \frac{\sigma_{2}'}{\sigma_{2}}\right) \nonumber \\
&= (1-d) (1-d\ind{crit}),
%(1-d)(1- \frac{E'}{E})
\label{damage}
\end{align}
where the superscript $'$ denotes all post-damage variables and where $d\ind{crit}$ is given by
\begin{equation}
d\ind{crit} = \
\min\left[ 1,  \frac{2 \tau_c}{(\mu^2 +1)^{1/2}-\mu} \frac{1}{\sigma_1 - q\sigma_2} \right],
\label{d_crit}
\end{equation} 
with $\tau_c$, the resistance to pure shear, or cohesion. 

%This approach differs from that taken in previous progressive damage models, for instance by ~\cite{Amitrano1999}, who prescribe an arbitrary, \textit{fixed} decrement in $d$ over each damaged element, and therefore use a \textit{static} sub-iteration for the evolution of damage to ensure that all local states of stresses lie within or on the failure enveloppe at the end of each loading time step. However, as 
We set the model time step equivalent to the time of damage propagation in the material, which we consider is lower bounded by the time of propagation of elastic waves that carry the damage information. The change in level of damage resulting from the stress redistribution between two neigbouring model elements distant by $\Delta x$ therefore occurs over an interval of time $\Delta t = \Delta x/c$, where $c$ is the speed of elastic waves. 
%slow loading and quasistatic conditions are ensured with our approach. 
%The mechanical behavior of the model in this limit is equivalent to that of ~\cite{Amitrano1999} and we verified that both models give the same results in terms of the angle of localization of the simulated damage.  

%---------------------------------------
%---------------------------------------

As done in previous progressive damage frameworks \citep[e.g.,][]{Tang1997, Zapperi1997, Amitrano1999}, disorder is introduced at the local scale in the failure criterion (i.e., the critical strength) via the cohesion parameter, $\tau_c$. The (adimensional) value of $\tau_c$ of a proportion $a$ of the model elements is randomly drawn from a uniform distribution, $[1-\epsilon, 1+\epsilon]$, and the cohesion of the remaining ($1-a$) of the elements is set to 1. Both control parameters on the distribution of $\tau_c$, $a$ and $\epsilon$, are varied. 

%Two types of disorder are also constrasted: time-independent, or ``quenched'', and time-dependant, or ``annealed''.
%In the later case, the cohesion of a model element is redrawn from the same initial distribution each time it is damaged. 

%\todo[inline]{Discuss here the properties of the disorder.}

Two types of disorder are considered: time-independent, or ``quenched'', and time-dependent, or ``annealed''.
In the case of quenched disorder, the local value of cohesion is set once, as an initial condition to the simulation, and does not evolve during the simulation. In the case of annealed disorder, the local value of cohesion is redrawn from a uniform distribution over $[1-\epsilon, 1+\epsilon]$ (i.e., $a$ is set to 1) everytime a model element gets damaged.
In the case of damage models, this approach is called ``annealed'' disorder \citep[e.g.,][]{Zapperi1997, Amitrano1999} and is thought of as to represent the heterogeneity in the strength of an elasto-brittle material during the damage increase.
Compared to quenched disorder, annealed disorder in such models does not affect the macroscopic mechanical behaviour nor the scaling laws characterizing the deformation of the material~\citep{HerrmannRoux1990}.
We note here that this terminology differs from the one used for classical elastic interfaces such as a fracture front:
\begin{itemize}
\item What we call here ``annealed'' disorder corresponds to a disorder that depends on the value of damage, which corresponds to the position of the interface, but not necessarily on time; hence, for a fracture front, this would be called ``quenched'' disorder.
\item What we call here ``quenched'' disorder is a disorder that does not depend on the position of the interface; for a fracture front, such disorder leads to the Larkin model~\cite{Larkin1970, Demery2014}, where the equation of evolution of the interface is linear.
\end{itemize}

\subsection{Simulation setup}
\label{setup}

\begin{figure}
\begin{center}
\includegraphics[width=8.3cm]{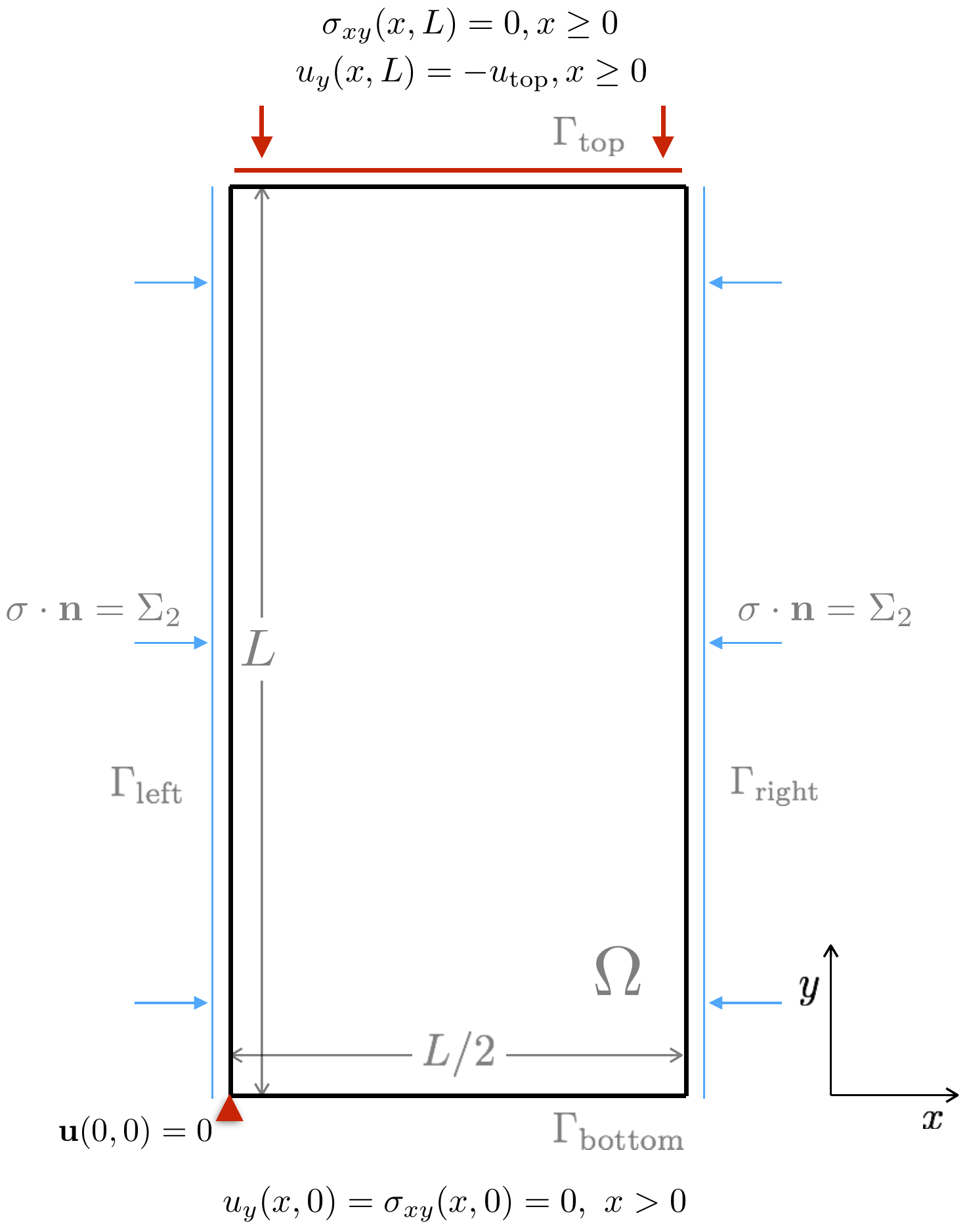}
\end{center}

\caption{Domain and boundary conditions for the uniaxial and biaxial compression simulations.} 
\label{Exp_setup}

\end{figure}

Uniaxial and biaxial compression simulations are conducted. 
The domain, $\Omega$, is rectangular with dimensions of $(L/2) \times L$ (Fig. \ref{Exp_setup}). The boundary, $\partial \Omega$, is partitioned as $\partial \Omega = \Gamma_{\text{top}} \cup \Gamma_{\text{left}} \cup \Gamma_{\text{bottom}} \cup \Gamma_{\text{right}}$. Compression is applied by prescribing a constant vertical velocity, $u_{\text{top}}$, on the top edge of the plate, $\Gamma_{\text{top}}$, with the opposite edge, $\Gamma_{\text{bottom}}$, maintained fixed in the direction of the forcing. 
In the uniaxial compression case, the lower left corner of the domain is maintained fixed in both the $x$ and $y$ directions and $\ssigma \cdot \mathbf{n} = 0$ on $\Gamma_{\text{left}} \cup \Gamma_{\text{right}}$. In the biaxial compression case, confinement is applied by prescribing a stress, $\Sigma_{2}$, on the lateral sides, $\Gamma_{\text{left}}$, $\Gamma_{\text{right}}$, such that the confinement ratio, $R  = \Sigma_{2}/\Sigma_{1}$, where $\Sigma_{1}$ is the stress integrated on the top side, is constant.

In all simulations, the prescribed confining stress and velocity on the top edge are both set small enough to ensure a small deformation regime. The total deformation of the domain (integrated on the top edge) after the formation of the macroscopic fault is of the order of $0.01\%$.
The ratio of the undammaged elastic modulus, $E^0$ and median value of cohesion, $\tau_c$, is chosen so that to be representative of a natural brittle material (rock or ice).
%The time for damage $\Delta x/c$, set by the speed of propagation of elastic waves in the material, $c$, and the spatial resolution of the model, $\Delta x$,
The time step $\Delta t$ is much smaller than the time $T = L/u_{\text{top}}$ associated with the deformation process. 
This large separation of scales between the rate of loading of the plate and the speed of evolution of damage in the material ensures quasi-static conditions. 

%---------------------------------------
%---------------------------------------

Finite elements and variational methods are used to solve the time-discretized problem on a Lagrangian grid within the C++ environment RHEOLEF~\citep{rheolef}. 
As cumulative deformations are small, the deformation of the mesh is not calculated and the position of grid nodes, not updated in time. Meshes with triangular elements are built using the Gmsh grid generator~\citep{Geuzaine2009}. 
In order to avoid introducing preferential orientations for the propagation of the damage, meshes are chosen unstructured. 
The average spatial resolution, $\Delta x$, is set by choosing the number $N$ of elements along the short side of the domain, such that $\Delta x = L/(2N)$. 
Except for the simulations comparing the effect of spatial resolution, $N$ is set to 80, and the mesh grid counts 33858 elements. 

All simulations are started from an initially undamaged material with uniform elastic modulus (except for the simulations shown in section~\ref{noise_on_E}) and stopped after the formation of a macroscopic fault (Fig. \ref{proj_hist}a).

Four parameters are varied in the simulations: (1) the local internal friction angle $\phi$, (2) the Poisson's ratio $\nu$, (3) the confinement ratio $R$ and (4) the type of disorder (quenched versus annealed). 
The range of values for each parameter is listed in Table \ref{parameters}. 
For each set of parameters, an ensemble of 25 simulations is run to estimate the average orientation of the fault. 
As the model is made adimensional with respect to the height $L$ of the domain, the median value of cohesion, $\tau_c$, and the velocity $u_{\text{top}}$ prescribed on the top edge, the results reported here are expressed in non-dimensional form.

%---------------------------------------
\begin{table}[h]
\centering

\renewcommand{\arraystretch}{1.5}

\begin{tabular}{   l  c  c  c}

\bf Parameters							& \bf 				& \bf Values 						\\ \hline 
\hline
Internal friction angle						& $\phi$			& $15^{\circ}, 30^{\circ}, 45^{\circ}, 60^{\circ}$				\\
Poisson's ratio							& $\nu$			& $0, 0.1, 0.2, 0.3, 0.4, 0.5$			\\
Confinement ratio						& $R$			& $0\%, 10\%, 20\%$			\\
Disorder								&				& quenched, annealed \\
%Number of elements along short edge (resolution)	& $N$		& 10, 20, 40, 80						\\
%Mean model resolution					& $\Delta x$		& $\frac{L}{2 N}, N =  10, 20, 40, 80$			\\
%Aspect ratio							& 				& 1, 2
\hline

\end{tabular}
\caption{List of parameters varied in the simulations and their range of values.}
\label{parameters}
\end{table}
%---------------------------------------

\subsection{Determination of the fault orientation}

The orientation, $\theta\ind{loc}$, of the simulated macroscopic faults is estimated using a projection histogram method. With this approach, the distribution histogram of a field value is calculated when projecting that field in a particular direction, $\beta$. By calculating projection histograms in all directions, the method allows detecting the principal orientations of linear features. The projection histograms of the field of damage are calculated as follow.

\begin{figure*}
\begin{center}
\includegraphics[angle = 0, scale=0.5]{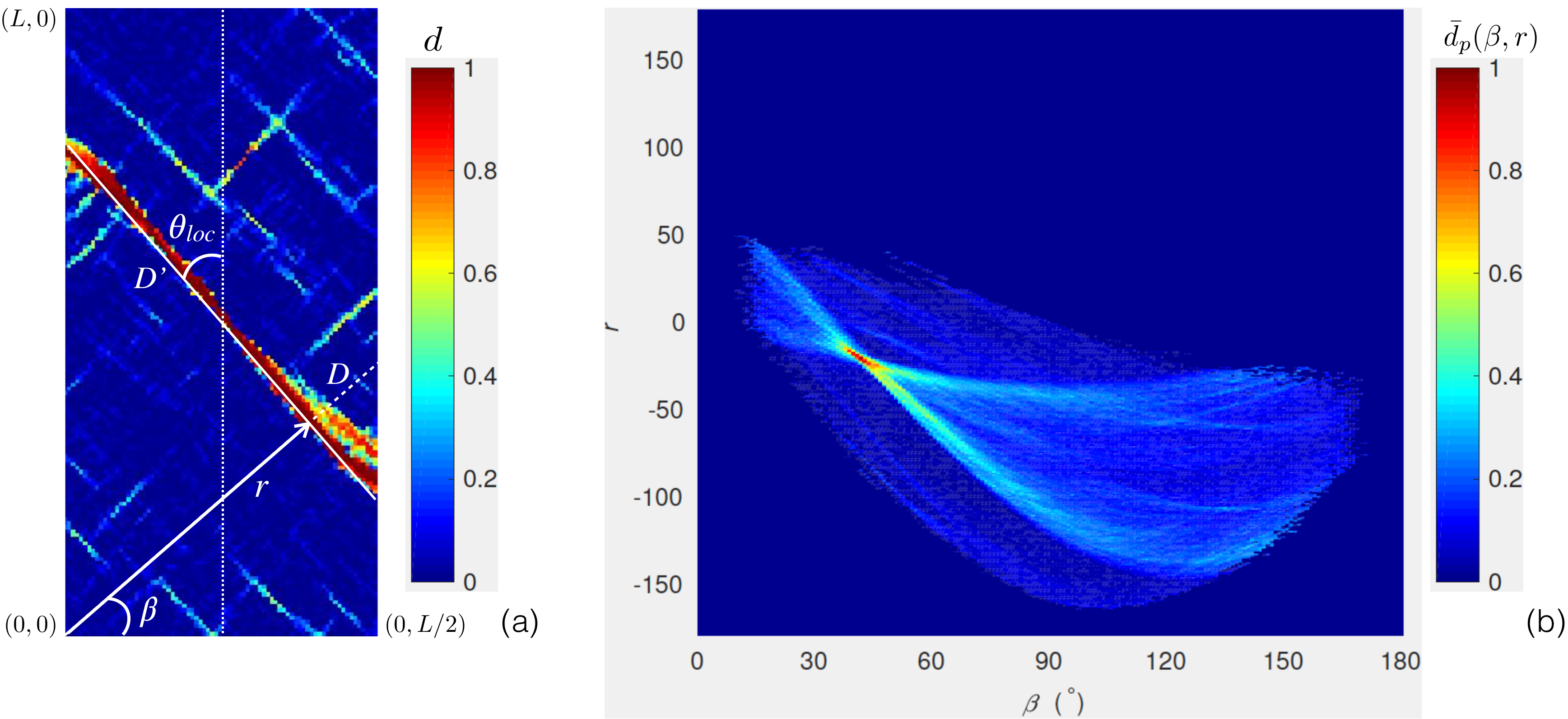}
\end{center}

\caption{(a) Field of the level of damage, $d$, interpolated onto a cartesian, square structured grid of $80 \times 160$ elements and (b) associated value of the projection histogram as a function of $r$ and $\beta$. The solid white line and white arrow on (a) represent the location, $r$, and orientation, $\theta\ind{loc}$, of the fault as estimated by the projection histogram method.} 
\label{proj_hist}

\end{figure*}

The instantaneous field of $d$ simulated on an unstructured grid is first interpolated onto a structured square elements grid of similar size ($N \times 2N$) using a nearest neighbor interpolation (Fig.~\ref{proj_hist}a). The origin of the rectangular image is defined as the lower left corner of coordinates $(x,y) = (0,0)$ and the direction, $\beta$, is defined relative to the axis $y = 0$. Hence the position of the center of any grid element $(x,y)$ can be written in polar coordinates as $(r \cos(\beta), r \sin(\beta))$, where $r=\sqrt{x^2+y^2}$ (Fig.~\ref{Exp_setup}). 

Any given direction $0^{\circ} \leq \beta \leq 180^{\circ}$ defines a line $D$ (dashed white line, Fig.~\ref{proj_hist}a) passing through the origin. For all positions $r$ along that line, the average level of damage of the grid elements found along the line $D'$ perpendicular to $D$ is calculated (solid white line, Fig.~\ref{proj_hist}a). 
The result, denoted $\bar{d}_p(\beta, r)$, is the projection histogram in the direction $\beta$.
The number of grid elements over which $\bar{d}_p(\beta,r)$ is calculated is not constant with $r$ and is smaller near the corners of the domain. Hence a minimum number of points is imposed as a threshold for the calculation of $\bar{d}_p(\beta, r)$, which corresponds to $N/4$. 

%Projections histograms are calculated for $0^{\circ} \leq \beta \leq 180^{\circ}$ and $-\N \leq r \leq N$. 
The normalized value of $\bar{d}_p$ corresponding to the instantaneous field of damage show in Fig.~\ref{proj_hist}a is plotted as a function of both $r$ and $\beta$ in Fig.~\ref{proj_hist}b. 

%---------------------------------------
%---------------------------------------

The localization angle, $\theta\ind{loc}$, is calculated using the absolute maximum value of the projection histogram for all values of $\beta$ and $r$, as 
\begin{eqnarray}
\theta\ind{loc} & = & \beta \ \text{ if } \beta < 90^{\circ} \\
& = & 180^{\circ} - \beta \ \text{ if } \beta > 90^{\circ} 
\end{eqnarray}
(Fig. \ref{proj_hist}a).
In the case of conjugate or multiple linear features, $\theta\ind{loc}$ corresponds to the orientation of the \textit{one} linear feature that returns the maximum in $\bar{d}_p$ (i.e., the most localized or most damaged feature).

We checked that (1) the model resolution, (2) the resolution of the regular square grid onto which the simulated fields of damage are interpolated, (3) the aspect ratio (square or rectangular domain) and (4) our choice of threshold for the minimum number of points for the calculation of $\bar{d}_p(\beta,r)$ have no effect on the results presented here.

\section{Numerical simulation results}
\label{sec:num_results}

\subsection{Mechanical behavior}
\label{mech}

\begin{figure}
\begin{center}
\includegraphics[width=8.3cm]{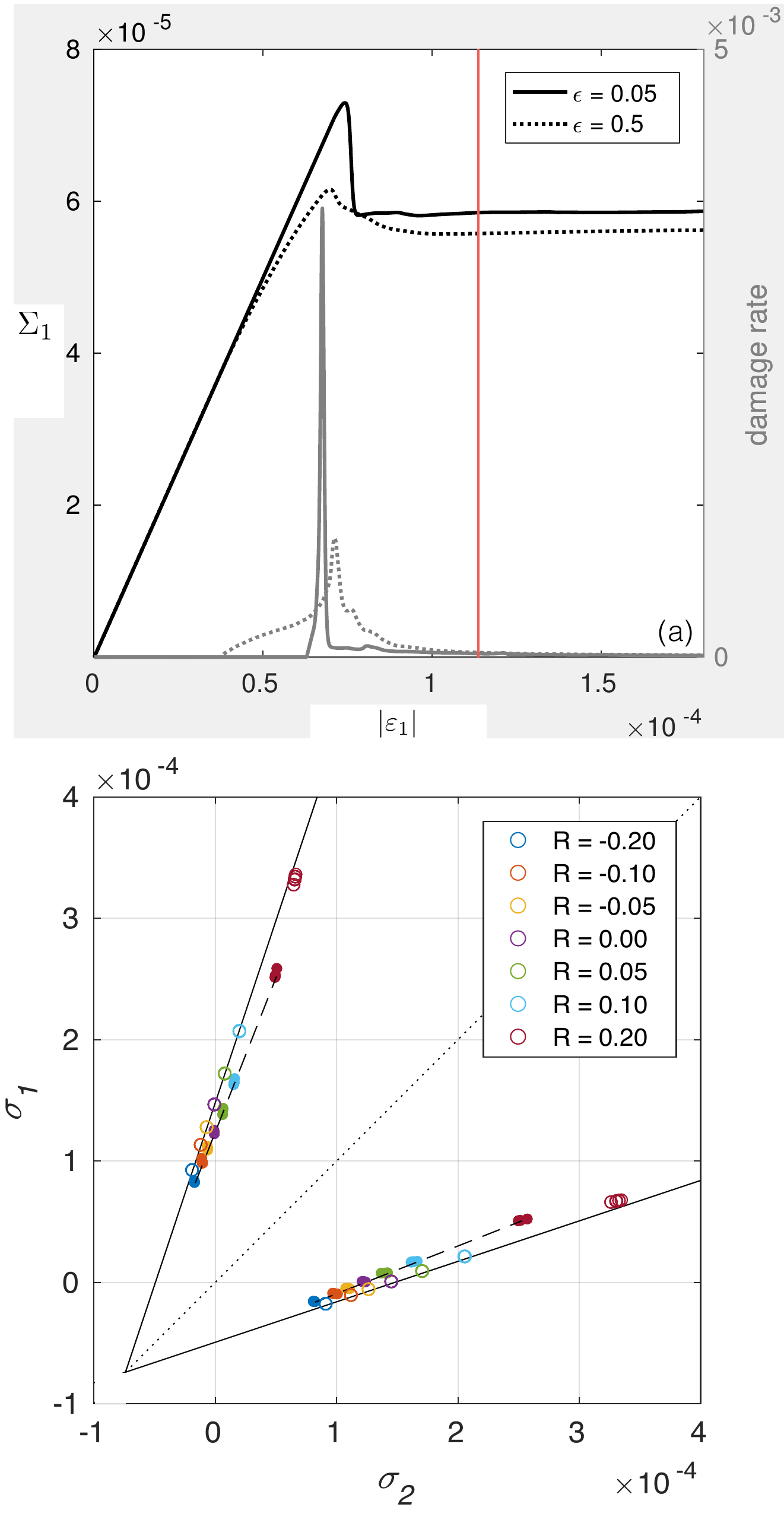}
\end{center}

\caption{(a) Macroscopic stress vs. macroscopic strain (black curve) and integrated damage rate (number of instantaneously damaged element times $1-d\ind{crit}$) in a typical compression simulation. The red line indicates the timing at which both the stress and structure of the fault stabilizes. (b) Macroscopic maximum and minimum principal stresses, $\Sigma_1$, $\Sigma_2$, (colored dots) estimated at the critical point before failure (maximum $\Sigma_1$) in a set of 5 simulations with $\phi = 30^{\circ}$ and using different confining ratios (biaxial compression for $R > 0$ and biaxial compression-tension for $R < 0$). The black solid lines represent the prescribed local MC criterion for $\tau_c = 1$ and $\phi = 30^{\circ}$. Cases of week ($\epsilon = 0.05$, $a = 1$, open circles) and strong ($\epsilon = 0.5$, $a = 1$, filled circles) are represented. Simulations with $\phi = 15^{\circ}$, $45^{\circ}$ and $60^{\circ}$ produced similar results. The dashed line represent a linear fit to the mean values of $\Sigma_1$ and $\Sigma_2$ estimated for the 5 simulations using $\epsilon = 0.5$.} 
\label{fig:mech_behavior}
\end{figure}

As a first step, we investigate the numerical simulations for the required brittle behavior in both a weak and strong disorder case. 
Fig.~\ref{fig:mech_behavior}a shows the macroscopic compressive stress, i.e., the normal stress integrated on the top boundary of the domain, $\Sigma_1$, as a function of the (prescribed) macroscopic normal strain on the top boundary, $|\epsilon_{1}|$ (black lines), together with the domain integrated damage rate, (the instantaneous number of damage elements times their distance to the damage criterion, $1 - d\ind{crit}$, grey lines) for two typical simulations using $\phi = 30^{\circ}$ and in which all elements have their cohesion, $\tau_c$, randomly drawn ($a = 1$) from a uniform distribution of width $\epsilon = 0.05$ (weak disorder, solid lines) and $\epsilon = 0.5$ (strong disorder, dashed lines). 
For weak disorder, the sharp drop in $\Sigma_1$ corresponds to the macrorupture and suggests that the simulated failure is indeed brittle. As expected from theory \citep{Kierfeld2006} and laboratory experiments \citep{Vasseur2015}, in the case of strong quenched disorder the propagation of damage in the model is slower and the drop in $\Sigma_1$ associated with the rupture smaller: the mechanical behavior appears therefore more ductile.
Consistent with the simulations of~\cite{Amitrano1999}, a less brittle behavior is also obtained by lowering the value of the prescribed internal friction angle $\phi$ (not shown). 
%As expected from theory \citep{Kierfeld2006} and laboratory experiments \citep{Vasseur2015}, in the case of very large quenched disorder, the propagation of damage in the model is slow (i.e., not brittle) and localization is impeded (Fig. \ref{fig:dam_fields}e). 

In both the weak and strong disorder cases, the damage rate indicates some precusory activity. Damage is initally distributed homogeneously over the domain (not shown). It localizes progressively as the loading is increased. After the macrorupture, both the damage (Fig.~\ref{proj_hist}a) and deformation (not shown) are highly concentrated along a fault or system of conjugate faults. 
Also in both cases, the macroscopic stress stabilizes after the macrorupture and damaging stops.
The field of $d$ at this point is analyzed via the projection histogram method to determine the angle of localization of the damage, $\theta\ind{loc}$.

As done in laboratory experiments on rocks~\cite{Byerlee1967, JaegerCook1979} and ice~\cite{Schulson2006, WeissSchulson2009}, we also verified that our simple model for the failure of a brittle material does reproduce a MC failure envelope at the macroscopic scale, i.e., the scale of the model domain.
The scattered plot on Fig. \ref{fig:mech_behavior}b represents the \textit{macroscopic} maximum and minimum principal stresses $\Sigma_1$ and $\Sigma_2$, at the point of rupture, defined at the maximum in $\Sigma_1$, in simulations with different levels of confinement (colored circles). 
For both weak (open circles) and strong (filled circles) disorder, the simulated macroscopic stresses at failure reproduce a MC enveloppe. 
It is important to note that the solid black line on Fig.~\ref{fig:mech_behavior}b represents the prescribed MC criterion for $\tau_c = 1$: the macroscopic states of stress at rupture do not lie exactly on this envelope due to the fact that lower values of $\tau_c$ are allowed in disorded cases.

\begin{figure*}

\begin{center}
\includegraphics[width=\linewidth]{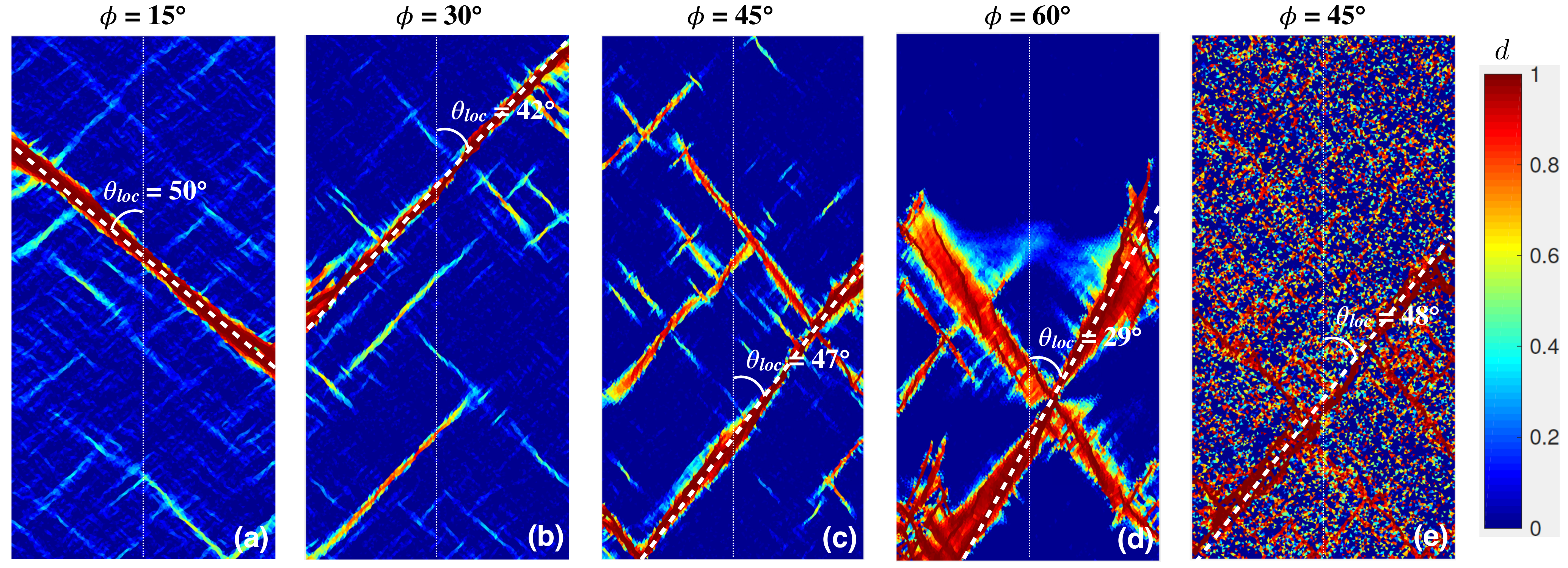}
\end{center}

\caption{Fields of the level of damage, $d$, simulated with $a = 1$, $\epsilon = 0.05$ and (a) $\phi = 15^{\circ}$, (b) $\phi = 30^{\circ}$, (c) $\phi = 45^{\circ}$, (d) $\phi = 60^{\circ}$. (e) Field of the level of damage simulated with $a = 1$, $\epsilon = 1$ and (a) $\phi = 45^{\circ}$.The white dashed line indicates the fault and the value of localization angle, $\theta\ind{loc}$, estimated by the projection histogram method in each case.} 
\label{fig:dam_fields}
\end{figure*}

In the weak disorder case, the slope, $q$, of the reproduced macroscopic enveloppe is, within about $5\%$, that prescribed at the element scale (solid black line on \ref{fig:mech_behavior}b, for $\tau_c = 1$). This suggests that the internal friction coefficient, $\mu$, in our model is scale-independant in the case of macroscopic brittleness. The stronger disorder case suggests a slope of the macroscopic enveloppe (Fig.~\ref{fig:mech_behavior}b, dashed black lines) that is $15\%$ lower than that prescribed at the element scale. This again is consistent with the ductile behavior of highly heterogeneous media~\citep{Kierfeld2006, Vasseur2015}. Also consistent with previous works on the brittle-ductile transition in the compressive failure of brittle materials, the discrepancy is larger for a higher lateral confinement~\citep{Renshaw2001} and lower values of $\phi$ (not shown)~\citep{Amitrano1999}. For $\epsilon = 1$, the present model approaches the limit of infinite disorder and localization is clearly reduced (Fig. \ref{fig:dam_fields}e). 
In this case, damage evolution becomes closer to a percolation problem~\citep{Roux1988}. 

In both the $\epsilon = 0.05$ and $\epsilon = 0.5$ cases, the differences between the microscopic and macroscopic value of $\mu$ will not affect the results of the simulations presented in the following, as these differences are much smaller than the discrepancies observed between the model, the Mohr-Coulomb and the linear stability analysis predictions. %which also increases with phi, the opposite of what is expected in the brittle-ductile transition. 
Nevertheless, to stay as close as possible to a truly brittle behavior, we restrict ourselves to $\phi \geq 15^{\circ}$ and $\epsilon \leq 0.5$ in the simulations analyzed here.

%---------------------------------------
%---------------------------------------

%---------------------------------------
%---------------------------------------

\subsection{Confinement ($R$) and dilatancy ($\nu$)}

We first perform a set of compression simulations representing a minimal disorder scenario. In this case, the field of cohesion is uniform everywhere except for a single element chosen at random. The value of $\tau_c$ for this element is initially set $0.05\%$ lower and is reset to the uniform value of its neighbors after the first damage event. 

\begin{figure}
\begin{center}
\includegraphics[width=8.3cm]{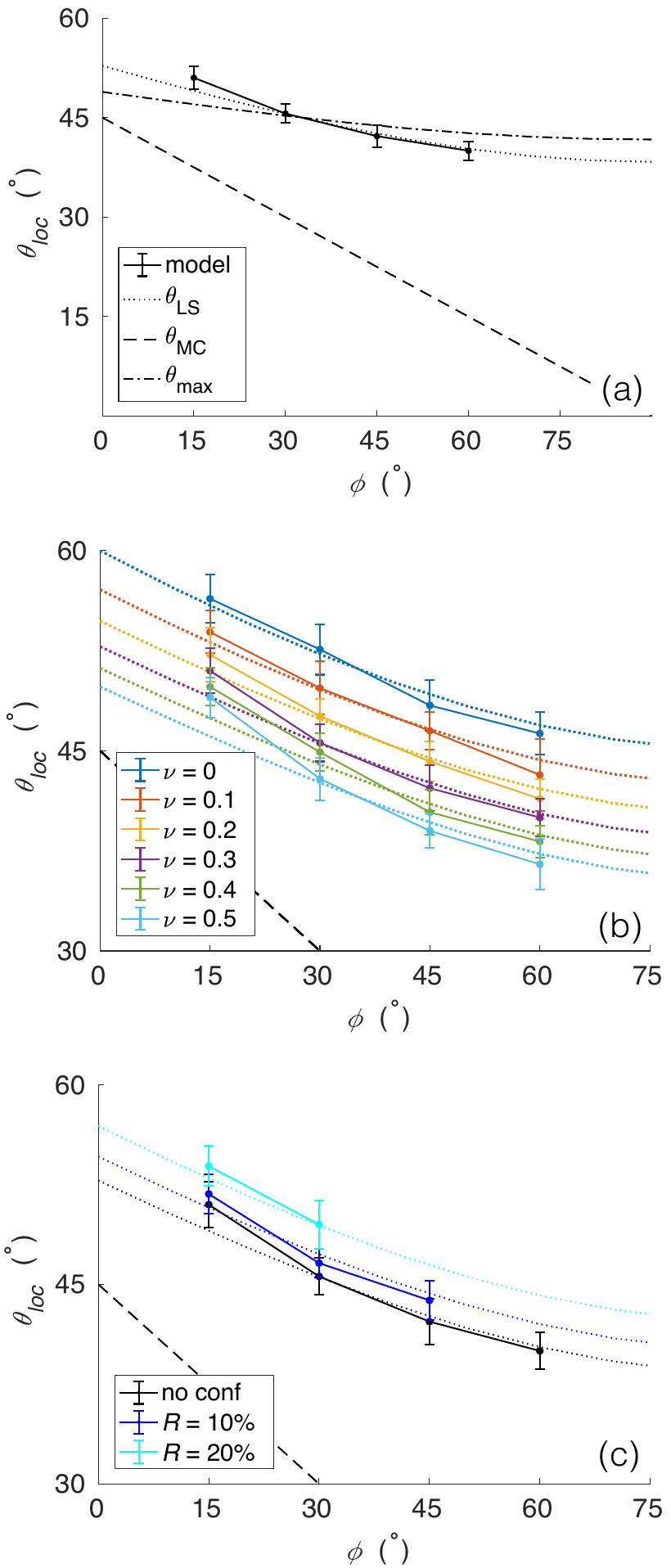}
\end{center}
\caption{(a) Mean $\theta\ind{loc}$ as a function of $\phi$ for an ensemble of simulations in which no confinement is applied and $\nu = 0$. The black dashed line shows to the MC prediction, $\theta\ind{MC}$, the dotted line, the angle of the most unstable mode, $\theta\ind{LS}$ and the dashed-dotted line, the angle of maximal MC stress redistribution, $\theta\ind{max}$. The error bars represent $\pm 1$ standard deviation from the mean.
Mean $\theta\ind{loc}$ for (a) different values of Poisson's ratio (no confinement) and (b) different values of confinement ratio ($\nu = 0.3$).The maximum confinement ratio, $R\ind{max}$, predicted by Eq. \eqref{eq:R_max} is of $58\%$ for $\phi= 15^{\circ}$, $33\%$ for $\phi = 30^{\circ}$, $17\%$ for $\phi = 45^{\circ}$ and of $7\%$ for $\phi = 60^{\circ}$.}
\label{fig:Eshelby}
\end{figure}

Fig.~\ref{fig:Eshelby}a shows the mean localization angle as a function of the internal friction angle $\phi$ calculated for an ensemble of 25 uniaxial compression (i.e, unconfined) simulations using identical boundary and loading conditions and started from different fields of cohesion, with $\nu = 0.3$.
The calculated $\theta\ind{loc}$ agrees well with the prediction from the linear stability analysis, $\theta\ind{LS}$ (dotted line).
The form and value of $\theta\ind{loc}(\phi)$ is far from both $\theta\ind{max}$ (dashed-dotted) and $\theta\ind{MC}$ (dashed line).
The agreement with $\theta\ind{LS}$ is kept when varying Poisson's ratio (Fig.~\ref{fig:Eshelby}b) and the confinement (Fig.~\ref{fig:Eshelby}c). 
%Fig.~\ref{fig:Eshelby} is also included in~\cite{short} (Fig. 2).

It is important to note that for $\phi = 15^{\circ}$, a systematic misfit with the linear stability analysis prediction is observed. In this case, a more ductile behavior is obtained. Some stress-shielding occurs adjacent to the fault and leads to a fault width larger than in the $\phi > 15^{\circ}$ cases (Fig. \ref{fig:dam_fields}a). 
Both this particular mode of failure and the potential error in the calculation of $\theta\ind{loc}$ associated with the width of the fault can explain this misfit.
On the other hand, for large $\phi$, boundary effects lead to a diffuse propagation of damage near the lateral boundaries, which explains the larger variability in the estimated value of $\theta\ind{loc}$ (Fig. \ref{fig:dam_fields}d). For this reason, the value of $\phi$ is limitted to $60^{\circ}$ in the simulations performed here. 

The simulations clearly show that $\theta\ind{loc}$ in our model depends on Poisson's ratio and on the degree of confinement. Both types of dependance are not accounted for by the MC theory. 
However, this minimum disorder scenario is very different than the way damage propagation is thought to occur in natural, disordered materials.
As there is only one incluson in a homogeneous matrix here and no other form of disorder introduced in the model, there is indeed no precursor to the rupture.

%----------------------------------
%----------------------------------

\subsection{Disorder}
\label{disorder}

\subsubsection{Effect of the width of the distribution of $\tau_c$ ($\epsilon$) and the initial number of inclusions ($a$)}

As a second step, we therefore address the effect of quenched disorder on the orientation of the simulated fault. Disorder can be augmented by either increasing the proportion $a$ of inclusions, i.e., of the number of model elements with a cohesion, $\tau_c$, drawn randomly from a uniform distribution, or by increasing the width, $\epsilon$, of this distribution of values of $\tau_c$. Here we explore the case of weak ($\epsilon = 0.05$) and strong ($\epsilon = 0.50$) disorder. 

\begin{figure}
\begin{center}
\includegraphics[width=8.3cm]{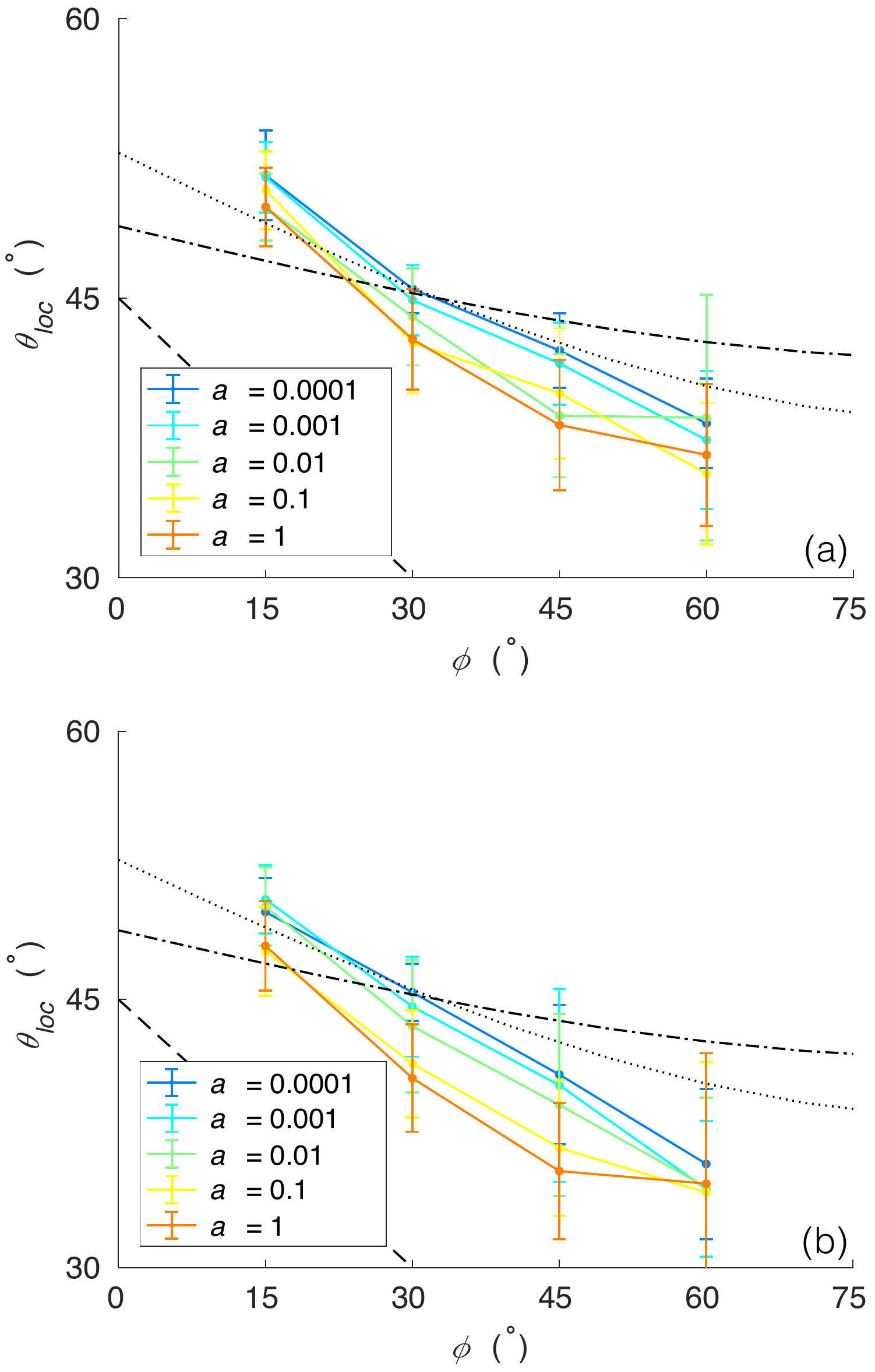}
\end{center}
\caption{$\theta\ind{loc}(\phi)$ in uniaxial (unconfined) experiments with $\nu=0.3$, different values of $a$ and (a) $\epsilon=0.05$ (weak disorder) and (b) $\epsilon=0.5$ (strong disorder).
%\red{The percentage difference between $\theta\ind{MC}$ and $\theta\ind{loc}$ varies between about x and x.
%}
}
\label{fig:disorder}
\end{figure}

Fig.~\ref{fig:disorder} shows the mean angle of localization of the damage as a function of $\phi$ for the two distribution widths and for different proportions of inclusions, $a$. 
A first observation is that disorder significantly affects the simulated fault orientation. The value of $\theta\ind{loc}$ departs from $\theta\ind{LS}$ as disorder is increased through either $a$ or $\epsilon$, even though the trend is qualitatively captured. 
In all cases, both the value and the form of the dependence of $\theta\ind{loc}$ on $\phi$ still do not agree with $\theta\ind{MC}$.
In both the weak and strong disorder scenarios, the form of the dependance of the orientation of the fault on Poisson's ratio and confinement is kept (see ~\cite{short}, Fig. 2c and 2d).%(Fig.~\ref{fig:disorder_nu_conf}). 

%----------------------------------

\begin{figure}

\begin{center}
\includegraphics[width=8.3cm]{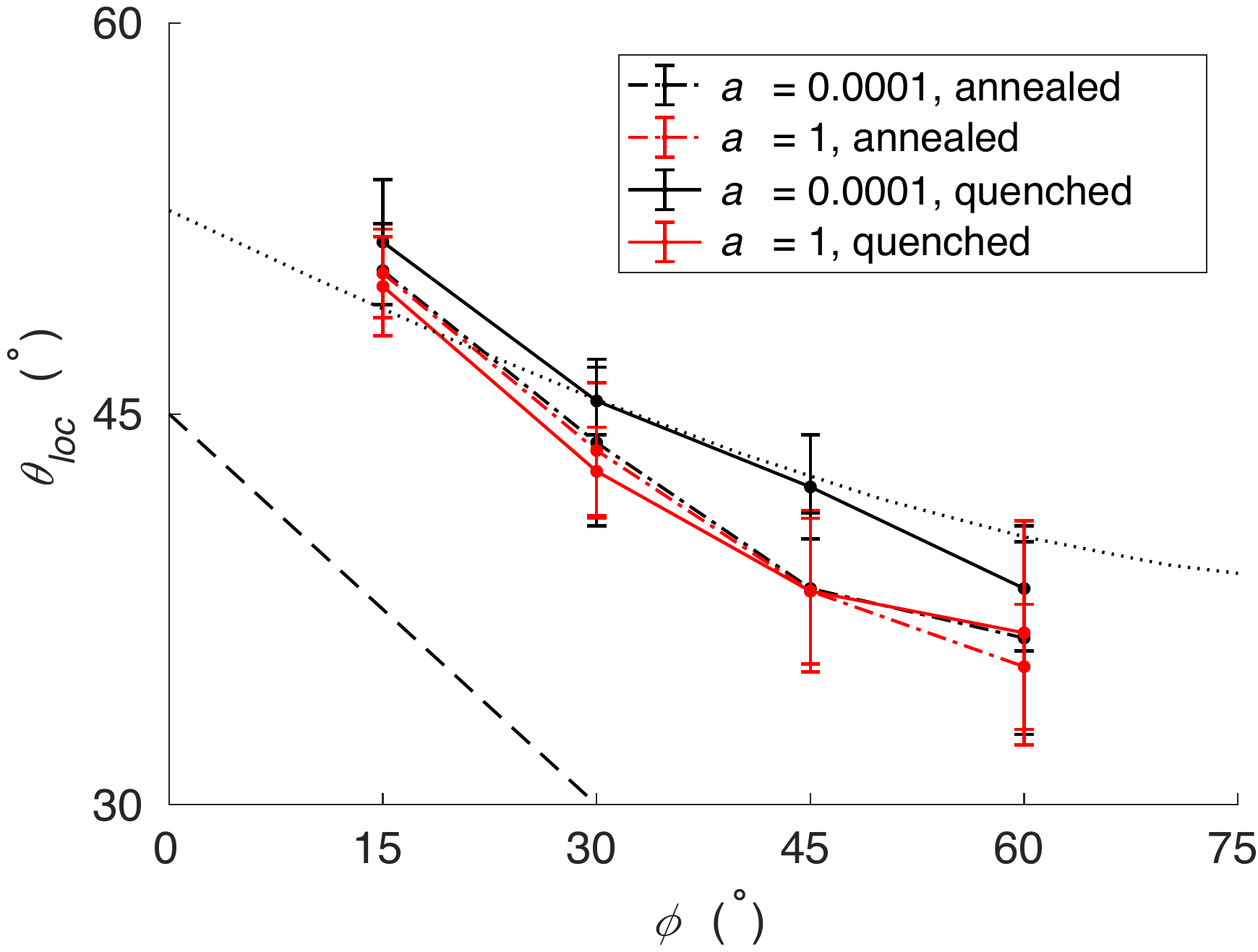}
\end{center}

\caption{$\theta\ind{loc}(\phi)$ for $\nu = 0.3$ no confinement and $\epsilon=0.05$, with $a = 0.0001$ (black lines) and $a = 1$ (red lines) and quenched (solid lines) versus annealed (dashed-dotted lines) disorder.} 
\label{fig:quenched_vs_annealed}
\end{figure}

We now compare the two types of disorder, quenched and annealed.
%Here, we investigate the effect of quenched and annealed disorder on the orientation of the macroscopic fault simulated with our damage model. 
Fig.~\ref{fig:quenched_vs_annealed} contrasts the estimated $\theta\ind{loc}$ using both types of disorder and a distribution of cohesion values of $\epsilon = 0.05$ for two widely different proportions of inclusions, $a = 0.0001$ and $a = 1$. 
The form of the dependance of $\theta\ind{loc}$ on $\phi$ remains similar using both approaches, but $\theta\ind{loc}$ departs further from the linear stability prediction in the case of few inclusions ($a = 0.0001$) with an annealed disorder. Considering our previous results on the effect of the level of disorder, this is expected: using an annealed disorder indeed increases the overall level of disorder, because the value of the cohesion of all damaged elements (not only the inclusions) becomes variable. The localization angle is therefore very sensitive to the type of disorder for few inclusions. Conversely, it is not sensitive to the type of disorder for $a = 1$.
%\todo{Laurent P.: Qualitatively, I agree: more disorder leads to more discrepancy between the theory and the simulations. However, we see that only one heterogeneities referred to as «quenched » is able to affect significantly the localization angle. What figure 8 might tell us is the following: a disorder that depends on the position of the interface  (in opposition to disorder that depends on x only) introduced non-linearities in the interface évolution (at least in standard depinning models), leading to a transient phase of progressive damage increase prior localization characterized by an avalanche-like dynamics. This complexity is not included in the stability analysis that assumes that localization takes place as soon as the first element damages. I have the feeling that as soon as there is a significant regime of progressive damage growth/interface propagation prior localisation, we do observe a discrepancy between theory and simulations. One single fixed heterogeneities is not enough to pin the crack and trigger a significant transient regime of damage growth. However, one single defect that is redrawn at each subsequent increment is able to do so.} \todo{VD:regarding the last sentence: I disagree, in the sense that this is not what is tested here. Here, each damaged element becomes a defect in the annealed case.}
Also, in the case of annealed disorder, the orientation of the fault does not appear sensitive to the initial number of inclusions (the results are the same for $a = 0.0001$ and $a = 1$), which suggests that disorder around the initial inclusions is not ``felt''.
What seems to count is the disorder of the elements within the path of the propagating fault. 
%\todo{L. P.: I would say instead that what counts is the presence of not a phase of progressive growth of damage prior localization. But I might be wrong. The best way to test this idea is to look at the evolution of the damage field prior localization for these different types of disorder (or even at the distribution of avalanche sizes if it is not too much work).}

%---------------------------------------
%---------------------------------------

\subsection{Noise on the Young modulus}
\label{noise_on_E}

We additionally test the effect of introducing noise in the field of the initial Young's modulus, $E^0$, instead of in the cohesive strength, $\tau_c$, on the fault localization angle. 
This later approach has also been used in progressive damage models, e.g., in~\cite{Amitrano1999}, leading to similar results in terms of the degree of localization of the damage, but is different on the point of view of the physics of disordered elastic systems
To make this difference clear, we can draw a parallel between the present damage model and elastic interfaces.
When disorder is introduced via $\tau_c$, the elastic interface evolves in a disordered medium, i.e., with a heterogeneous mechanical strength. 
Conversely, incorporating disorder in the field of $E$ amounts to introducing disorder via the initial field of damage, as $E = E^0(1-d)$. The elastic interface then starts from a deformed configuration, but evolves within a homogeneous medium. 

\begin{figure}

\begin{center}
\includegraphics[width=8.3cm]{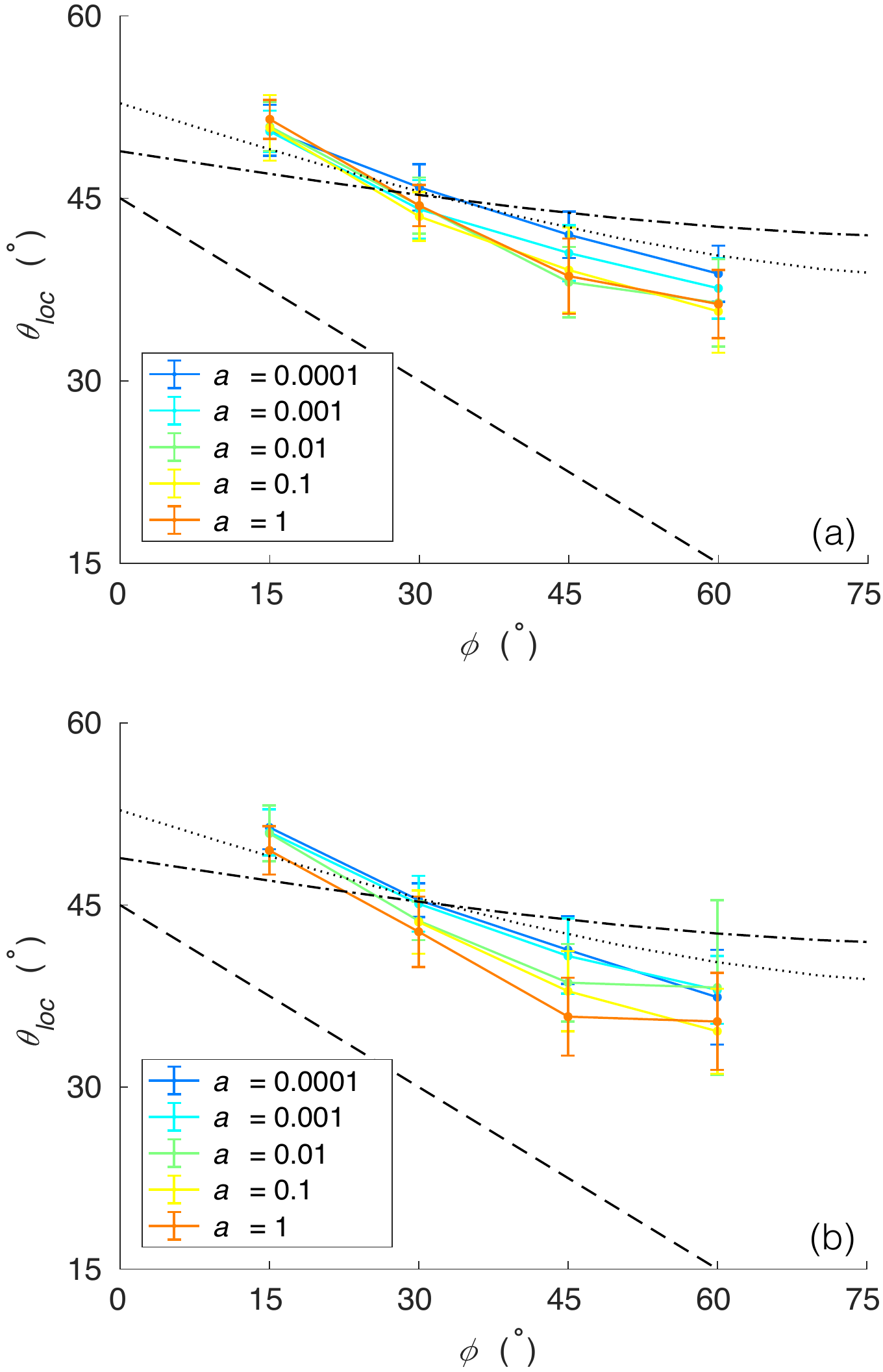}
\end{center}

\caption{$\theta\ind{loc}(\phi)$ for $\nu = 0.3$, $\epsilon=0.05$ and different $a$, in the case of quenched disorder introduced in (a) the initial field of cohesion, $\tau_c$, and (b) the initial Young's modulus, $E^0$.} 
\label{fig:E0_vs_C}
\end{figure}

Simulations in which quenched disorder is introduced in the initial field of $E$ by drawing the value of $E^0$ from the same distribution as used for the field of $\tau_c$ indicate that the results are not affected by this distinction (Fig.~\ref{fig:E0_vs_C}).
%, at least with respect to the formation of a localized, macroscopic fault and its orientation.
The form and value of $\theta\ind{loc}(\phi)$ are indeed similar to that obtained when disorder is introduced via the field of $\tau_c$ (Fig. \ref{fig:disorder}), in both the weak ($\epsilon = 0.05$, Fig. \ref{fig:E0_vs_C}a) and strong ($\epsilon = 0.50$, Fig. \ref{fig:E0_vs_C}b) disorder cases and the disagreement with $\theta\ind{MC}$ or $\theta\ind{max}$ is still clear.

This behavior strongly differs from the behavior of elastic interfaces with a negative kernel, where the interface always evolves towards a flat configuration without disorder, irrespective of its initial shape. This is reminiscent of the differences between damage models and the classical depinning problem stressed in Sec.~\ref{sec:intro}.

%---------------------------------------
%---------------------------------------

%%---------------------------------------
%\begin{figure*}
%
%\begin{center}
%\includegraphics[width=16cm]{E0_annealed}
%\end{center}
%
%\caption{$\theta\ind{loc}(\phi)$ for $\nu = 0.3$, $\epsilon=0.05$ and different $a$, in the case of quenched and annealed disorder introduced in the initial Young's modulus, $E^0$, for the case of (a) weak ($\epsilon = 0.05$) and (b) strong ($\epsilon = 0.50$) disorder.} 
%\label{E0_annealed}
%%
%\end{figure*}
%%---------------------------------------

\subsection{Summary of results}

We summarize our main findings from the numerical simulations below.  

%%---------------------------------------
%\begin{figure}
%\centering
%
%\includegraphics[width=8.3cm]{collapse}
%
%\caption{$\theta\ind{loc}(\phi)$ for quenched and annealed disorder simulations in which disorder is introduced via $\tau_c$ or $E$, with $\nu \geq 0.3$, $\epsilon \in [0.05, 0.5, 1]$ and $a \in [0.0001, 0.001, 0.01, 0.1, 1]$.} 
%\label{fig:collapse}
%%
%\end{figure}
%%---------------------------------------

\begin{enumerate}

\item With the Mohr-Coulomb criterion prescribed at the local scale, our simple progressive damage, linear-elastic model reproduces the macroscopic Mohr-Coulomb enveloppe (Fig.~\ref{fig:mech_behavior}b). 
However, the macroscopic angle of localization of damage, $\theta\ind{loc}$ does not correspond to the prescribed microscropic internal angle of friction, $\phi$, predicted by Coulomb's theory of failure. It does not correspond to the angle of maximum redistribution of the stress either. % (Fig.~\ref{fig:Eshelby}a). 

\item The agreement with the angle of localization predicted by the spatial dependence of the elastic kernel of interactions established by the linear stability analysis of the damage model is best in the case of the minimal disorder simulations. %(Fig.~\ref{fig:Eshelby}). 
This particular scenario is however far from damage propagation in real materials (no precursor to the rupture). It should therefore be interpreted as a theoretical rather than a physical limit. 

\item $\theta\ind{loc}$ is sensitive to dilatancy (Poisson's ratio) and to confinement. The form of the dependance to both parameters is in agreement with that predicted by the linear stability analysis. 
%(Fig.~\ref{fig:Eshelby}b, c). 
This dependency is not accounted for by the MC failure theory. 

\item The localization angle diverges from the linear stability analysis prediction as soon as disorder becomes more important: by adding more inclusions, using a larger distribution for the mechanical strength of the inclusions or by using an annealed disorder. 

\item Introducing disorder via the elastic modulus ($E^0$) instead of the critical strength (i.e., $\tau_c$) does not affect our results.
%That localization occurs in our damage model when disorder is introduced via the elastic modulus while other mecanical parameters are homogeneous is incompatible with a negative kernel.

\end{enumerate}

\section{Conclusion}
\label{sec:conclusion}

We have computed the elastic kernel associated with a general progressive damage model.
We have applied our general result to a specific model of failure under compression, which implements the Mohr-Coulomb failure criterion at the local scale and has been used to model the failure of heterogeneous solids such as rock and ice, and we have shown that the elastic kernel has unstable modes.
In this sense, the field of damage in a damage model is thus very different from the height of an elastic interface~\cite{Kardar1998} or even the field of accumulated plastic strain in a sheared amorphous solid~\cite{Lin2014b}.

We have associated the presence of unstable modes in the elastic kernel to the localization of damage along a macroscopic fault which is observed in numerical simulations~\cite{Amitrano1999} and in experiments~\cite{Lockner1991}.
We have tested this relation in numerical simulations.
With small disorder, we have shown that the orientation of the fault is indeed given by the most unstable mode of the elastic kernel.
Increasing the disorder, the orientation of the fault deviates from the orientation of the most unstable modes, which suggests that localization results from a complex interaction between the unstable modes of the kernel and the disorder.

In order to understand the evolution of damage in numerical simulations or experiments, one thus has to understand the behavior of an interface with unstable modes in an heterogeneous environment, which is a formidable challenge.

\begin{acknowledgments}
V. Dansereau is supported financially by TOTAL S.A.. E. Berthier has been supported by the program Emergence from UPMC.
We thank D. Kondo and A. Rosso for usefull discussions and comments and A. Amon for providing the code for the projection histogram calculations.
\end{acknowledgments}

\end{document}